\documentclass[a4paper,12pt]{article}
\usepackage{jheppub,booktabs}
\usepackage{amsmath,amssymb,slashed}
\usepackage{braket}
\usepackage{xcolor,colortbl}
\usepackage{multirow}
\usepackage{cancel}
\usepackage{subfigure}
\usepackage{mathrsfs}
\usepackage{graphicx}
\usepackage[compat=1.1.0]{tikz-feynman}
\usepackage{pdflscape}
\usepackage{feynmf}
\usepackage{enumerate}
\usepackage[utf8x]{inputenc}
\usepackage{xspace}
\usepackage[normalem]{ulem}

\newcommand{\dd}{\mathrm{d}}

\newcommand{\eV}{\ \mathrm{eV}}

\newcommand{\GeV}{\ \mathrm{GeV}}
\newcommand{\TeV}{\ \mathrm{TeV}}
\newcommand{\fb}{\ \mathrm{fb}}
\newcommand{\ab}{\ \mathrm{ab}}

\newcommand{\kpc}{\ \mathrm{kpc}}
\newcommand{\cm}{\ \mathrm{cm}}
\newcommand{\second}{\ \mathrm{s}}

\newcommand{\lv}{\left\vert}
\newcommand{\rv}{\right\vert}
\newcommand{\micrOMEGAs}{\texttt{micrOMEGAs}\xspace}
\newcommand{\sigmav}{\braket{\sigma v}}

\def\lrvert#1{{\left\vert #1 \right\vert}}

\author[a]{Subhaditya Bhattacharya,}
\author[b]{Sven Fabian,}
\author[b]{Johannes Herms,}
\author[b]{Sudip Jana}
\affiliation[a]{Department of Physics, Indian Institute of Technology, Guwahati, Assam 781039, India}
\affiliation[b]{Max-Planck-Institut f{\"u}r Kernphysik\\ Saupfercheckweg 1, 69117 Heidelberg, Germany}
\emailAdd{subhab@iitg.ac.in, sven.fabian@mpi-hd.mpg.de, herms@mpi-hd.mpg.de, sudip.jana@mpi-hd.mpg.de}

\title{Flavor-Specific Dark Matter Signatures through the Lens of Neutrino Oscillations}

\abstract{We investigate the flavor-specific properties of leptophilic dark matter in neutrino mass models, where dark matter signals are directly correlated with the neutrino oscillation data, providing complementary insights into the neutrino mass hierarchy and CP phases. Notably, this can be accomplished without introducing a flavor-specific portal to dark matter, imposing any new flavor symmetry, or involving flavon fields. As a case study, we analyze the correlation between the flavor-philic nature of dark matter and neutrino oscillation data in the type-II seesaw and Zee-Babu models, and extend this discussion to other neutrino mass models. We analyze the indirect signatures of such leptophilic dark matter, specifically examining the spectrum of the cosmic ray electron/positron flux resulting from the pair annihilation of dark matter in the Galactic halo, and explore correlated lepton-specific signals at collider experiments sensitive to neutrino oscillation data.
}

\begin{document}
\maketitle

\clearpage

\section{Introduction}
The compelling evidence supporting the existence of dark matter (DM) in the Universe serves as a robust call for physics beyond the Standard Model (BSM). In spite of the great experimental progress in the search for non-gravitational signatures of DM in the laboratory and the cosmos, we still know little about its nature: its non-gravitational interactions have not yet been established, while deviations of its behavior from that of cosmological dust have been tightly constrained. Null results put increasing pressure on the most plausible scenarios for DM. One of these is \textit{Weakly Interacting Massive Particle} (WIMP) DM, whose production mechanism links the precisely measured DM relic abundance~\cite{Planck:2018vyg} inversely to the DM annihilation cross section in the early Universe, making concrete particle physics predictions. This paper investigates the possibility where the DM annihilation signature is linked to a second open question in particle physics: the origin of neutrino masses and mixing. 

%%%%%%%%%%%%%%%%%%%%%%%%%%%%%%%%%%%%%%%%%%%%%%%%%%%%%%%
\begin{figure}[htb!]
\centering
\begin{tikzpicture}
\begin{feynman}
\vertex (d1) {};
\vertex[below=1.cm of d1] (d2) {};
\vertex[below=1.cm of d2] (d3) {};
\vertex[below=1.cm of d3] (d4) {};

\vertex at ($(d1)!0.5!(d2) - (8cm,0)$) (a1) {\(\mathrm{DM}\)};
\vertex at ($(d3)!0.5!(d4) - (8cm,0)$) (a2) {\(\mathrm{DM}\)};

\vertex at ($(d1)!0.5!(d2) - (2cm,0)$) (c1);
\vertex at ($(d3)!0.5!(d4) - (2cm,0)$) (c2);

\vertex at ($(c1)!0.5!(c2) - (4cm,0)$) (b1);

\diagram* {
(a1) -- [solid] (b1),
(a2) -- [solid] (b1),

(b1) -- [scalar] (c1),
(b1) -- [scalar] (c2),

(d1) -- [solid, with arrow=0.3] (c1),
(d2) -- [solid, with arrow=0.3] (c1),
(c2) -- [solid, with arrow=0.7] (d3),
(c2) -- [solid, with arrow=0.7] (d4)
};
\end{feynman}
\filldraw[fill=white, draw=white] (b1) ellipse (0.5cm and 0.5cm);
\filldraw[pattern=north west lines, pattern color=blue, draw=black] (b1) ellipse (0.5cm and 0.5cm);
\filldraw[fill=blue!20!white, draw=black] (c1) ellipse (0.7cm and 0.5cm) node[anchor=center]{\color{red!80!black} $\left( m_{\nu} \right)_{ij}$};
\filldraw[fill=blue!20!white, draw=black] (c2) ellipse (0.7cm and 0.5cm) node[anchor=center]{\color{red!80!black} $\left( m_{\nu} \right)_{ij}$};
\node[] at (0.2,0) {$\ell_{i}^{+}$};
\node[] at (0.6,-1) {$\ell_{j}^{+} \left( \overline{\nu}_{j} \right)$};
\node[] at (0.2,-2) {$\ell_{i}^{-}$};
\node[] at (0.6,-3) {$\ell_{j}^{-} \left( \nu_{j} \right)$};
\node[rotate=14.5] at (-4.1,-0.75) {$\mathcal{S}^{++} \left( \mathcal{S}^{+} \right)$};
\node[rotate=-14.5] at (-4.1,-2.25) {$\mathcal{S}^{--} \left( \mathcal{S}^{-} \right)$};
\end{tikzpicture}
\caption{Schematic diagram for Dark Matter annihilations to SM leptons via generic, doubly (singly) charged scalars~$\mathcal{S}^{\pm\pm} \left( \mathcal{S}^{\pm} \right)$.}
\label{Fig:SchematicAnnihilation}
\end{figure}
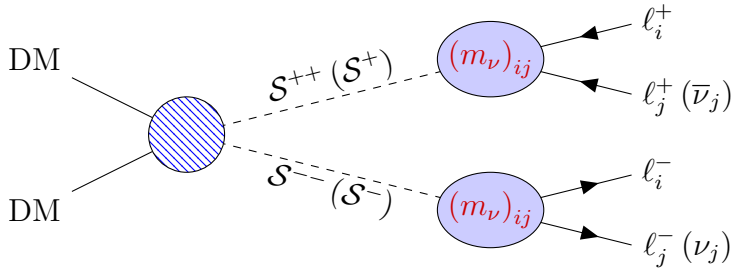
%%%%%%%%%%%%%%%%%%%%%%%%%%%%%%%%%%%%%%%%%%%%%%%%%%%%%%%

We primarily focus on purely scalar extensions of the SM, where a SM-singlet scalar WIMP~$S$, stabilized by some symmetry, annihilates into SM particles through terms in the scalar potential. In its simplest realization, the DM phenomenology is completely determined by its interaction through the \textit{Higgs-portal}, $V \supset H^\dagger H S^2$, which allows annihilation of a pair of DM particles into SM final states via Higgs mediation. However, this coupling also induces large DM-nucleon scattering cross sections observable in DM direct detection (DMDD) experiments.  Given the null observation in DMDD, Higgs-portal scalar WIMP is mostly ruled out, except for the Higgs resonance region (i.e., $m_\mathrm{DM} \sim 60 \GeV$) and a narrow range around 3 TeV~\cite{Cirelli:2024ssz}, considering perturbative limits.\footnote{Note here, DMDD experiments have limited sensitivity in the Higgs resonance region, but the narrow region near 3 TeV could be entirely excluded by future DMDD searches.} This conclusion, however, changes once other BSM scalars are taken into account. In particular, BSM physics to address neutrino masses may play a crucial role in DM phenomenology. 

Amongst neutrino mass generation mechanisms, the type-I seesaw scenario requires extension in the fermion sector~\cite{Minkowski:1977sc, Mohapatra:1979ia, Yanagida:1979as, Gell-Mann:1979vob, Glashow:1979nm}. Such BSM fermions may either themselves be ``sterile neutrino dark matter''~\cite{Dodelson:1993je, Shi:1998km}, or constitute the so-called ``neutrino-portal''~\cite{Falkowski:2009yz, Lindner:2010rr, Gonzalez-Macias:2016vxy} that may mediate between DM and the SM. However, a purely scalar extension of the SM can give rise to neutrino masses just as well. The simplest instance is the type-II seesaw scenario~\cite{Schechter:1980gr, Cheng:1980qt, Mohapatra:1980yp, Lazarides:1980nt}, in which an electroweak triplet scalar is added to the SM. Its most distinctive feature, perhaps, is the existence of a doubly charged scalar in the spectrum. Doubly charged scalars can also participate in neutrino mass generation in more complex or loop-suppressed scenarios. Such additional scalars can act as portals mediating between scalar DM and the SM thermal bath in the early Universe through the scalar potential. Such \emph{scalar-neutrino-portal DM} scenarios often lead to leptophilic DM, which is less affected by the DMDD constraints, due to loop suppressed interactions~\cite{Barman:2021hhg}. Importantly, the annihilation products\footnote{The dark matter candidate can annihilate into neutrinos, but this channel is less relevant for detection, as it also annihilates into charged leptons. Notable sensitivity differences between these channels are highlighted in Ref.~\cite{Cirelli:2024ssz}.} of such DM feature a specific flavor structure that is linked to neutrino oscillation observables, as schematically depicted in Fig.~\ref{Fig:SchematicAnnihilation}.

The scenario leads to flavor-specific DM signatures, without extending the gauge symmetry, or introducing additional flavor symmetries or involving multiple flavon fields.

In this paper, we investigate the link between neutrino oscillation observables and flavored DM signatures in the scenario where a doubly charged scalar is responsible for neutrino mass generation and DM annihilation.
Flavored signatures can appear both in DM indirect detection and in collider searches.

In DM indirect detection (DMID), the couplings of the mediator to the lepton flavors determine the spectrum of cosmic ray positrons expected from DM annihilation in our Galaxy.
This has received attention in the light of various cosmic ray excesses observed over the past decades (see in particular Refs.~\cite{Gogoladze:2009gi,Dev:2013hka,Li:2017tmd,Li:2018abw,Sui:2017qra}). Notably, the rise of positron/electron flux ratio towards TeV energies was first observed by PAMELA~\cite{Adriani:2008zq} and ATIC~\cite{Chang:2008aa} (which may plausibly be of astrophysical origin, while DM explanations are subject to strong gamma ray constraints, see e.g.~\cite{Cirelli:2024ssz,Lin:2014vja,Xiang:2017jou}), while a more recent but inconclusive excess in the electron plus positron spectrum has been observed by DAMPE~\cite{DAMPE:2017fbg}.

Even though its detailed origin is uncertain, the smooth positron flux observed by AMS-02~\cite{AMS:2021nhj} has been used to place stringent bounds on leptophilic DM from the absence of sharp spectral features above a smooth background flux~\cite{John:2021ugy}.
Future experiments like the AMS-100 proposal~\cite{Schael:2019lvx} may greatly increase the reach of DM searches via positrons towards $\sim$TeV scale masses.

In the present work, we explore the range of scalar-neutrino-portal DMID signals that is compatible with neutrino oscillation data.
We investigate the possibility of discerning the neutrino mass ordering through DM searches and assess whether the neutrino Majorana phases, to which neutrino oscillation experiments are insensitive, can be linked to DMID sigatures in the present scenario. We investigate collider signatures of the doubly charged scalars in a similar spirit.

The paper is organized as follows: initially, we provide a concise overview of neutrino mass models, focusing on establishing a direct correlation between leptophilic DM and observable neutrino oscillations. Following that, we analyze the phenomenology of DM, examining the relic abundance constraints. Subsequently, we conduct a detailed analysis of DM indirect detection and its flavor dependence. Additionally, we explore the complementarity of collider experiments in the context of flavor-specific scenarios. Finally, we present our conclusions.

\section{Neutrino mass models and charged-scalar-portal dark matter}
Understanding the tiny masses of neutrinos is crucial for explaining the observed neutrino oscillations (cf. Tab.~\ref{Tab:NeutrinoOscillationData}).
%%%%%%%%%%%%%%%%%%%%%%%%%%%%%%%%%%%%%%%%%%%%%%%%%%%%%%%
\begin{table}[b!]
    \centering
    \begin{tabular}{cccccc}
        \toprule
        Parameter & Normal Hierarchy (NH) & Inverted Hierarchy (IH)\\
        \midrule
        $\sin^{2}\theta_{12}$ & $0.304$ & $0.304$ \\
        $\sin^{2}\theta_{23}$ & $0.573$ & $0.575$ \\
        $\sin^{2}\theta_{13}$ & $0.02219$ & $0.02238$ \\
        $\delta_{\mathrm{CP}}/{}^{\circ}$ & $197_{-24}^{+27}$ & $282_{-30}^{+26}$ \\
        $\Delta m_{21}^{2}/\left( 10^{-5} \eV^{2} \right)$ & $7.42$ & $7.42$ \\
        $\Delta m_{3\ell}^{2}/\left({10^{-3}} \eV^{2}\right)$ & $2.517$ & $-2.498$ \\
    \hline
  \end{tabular}
    \caption{Summary of the neutrino oscillation parameters with Super-Kamiokande atmospheric neutrino data~\cite{Esteban:2020cvm}. The difference between $m_{3}^{2}$ and $m_{\ell}^{2}$ depends on the mass hierarchy ($\ell=1(2)$ for NH(IH)).}
    \label{Tab:NeutrinoOscillationData}
\end{table}
%%%%%%%%%%%%%%%%%%%%%%%%%%%%%%%%%%%%%%%%%%%%%%%%%%%%%%%%%%%%%%%%%%%%%%%%%%%%%%%%%%%%%%%%%%%%%%%%%%%%%
Many theories propose that neutrinos, assumed to be Majorana particles, obtain their masses from higher-dimensional operators; the lowest being the dimension-five operator $\mathcal{O}_1=L^i L^j H^k H^l \epsilon_{i k} \epsilon_{j l}/\Lambda$, suppressed by a high-energy scale ($\Lambda$) linked to lepton number violating BSM physics~\cite{Weinberg:1979sa}. In the above, $L$ represents the lepton doublet and $H$ the SM Higgs doublet with the $SU(2)_{L}$ indices~$i,j,k,l$. The seesaw mechanism exemplifies such kind of BSM scenarios. Phenomenologically, greater interest and testability arises when the scale of lepton number violation lies proximate to the electroweak scale, typically within the sub-TeV or TeV range, so that it can be probed in collider and other experiments. The required suppression can be attributed to loop factors, higher-dimensional operators, or a combination thereof. While many neutrino mass models rely on matter field extensions and the incorporation of right-handed neutrinos, our focus is directed toward models where minimal scalar field extensions alone suffice to generate neutrino masses and mixings. Specifically, we investigate scenarios wherein a doubly charged scalar emerges.

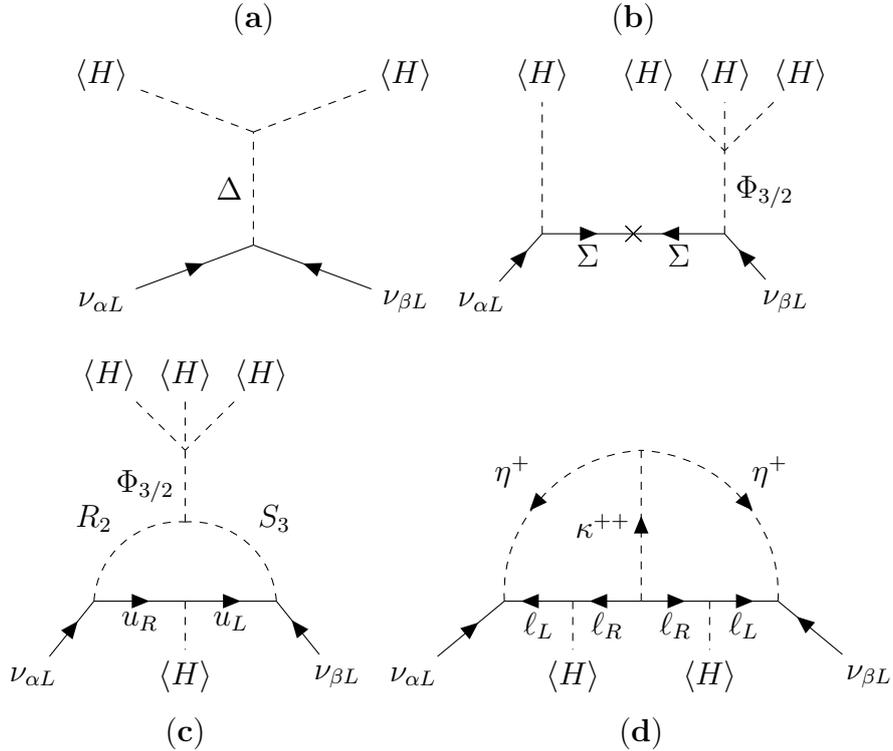
\begin{figure}[b!]
\centering
\begin{tikzpicture}
\node[] at (2cm,3.75cm) {$\mathbf{(a)}$};
\node[] at (7cm,3.75cm) {$\mathbf{(b)}$};
\begin{feynman}
%Class 1
\vertex (a1) {\(\nu_{\alpha L}\)};
\vertex[right=4cm of a1] (a2) {\(\nu_{\beta L}\)};
\vertex[above=3cm of a1] (e1) {\(\braket{H}\)};
\vertex[above=3cm of a2] (e2) {\(\braket{H}\)};

\vertex at ($(a1)!0.5!(a2) + (a1)!0.25!(e1)$) (b1);
\vertex at ($(a1)!0.5!(a2) + (e1)!0.25!(a1)$) (d1);

%Class 2
\vertex[right=1cm of a2] (a3) {\(\nu_{\alpha L}\)};
\vertex at ($(a3) + (a2) - (a1)$) (a4) {\(\nu_{\beta L}\)};
\vertex at ($(a3)!0.2!(a4) + (a1)!0.3!(e1)$) (b2);
\vertex at ($(a3)!0.8!(a4) + (a1)!0.3!(e1)$) (b4);
\vertex at ($(b2)!0.5!(b4)$) (b3);

\vertex at ($(e2) + (a3) - (a2) + (a3)!0.2!(a4) - (a3)$) (d2) {\(\braket{H}\)};
\vertex at ($(d2) + (b4) - (b2)$) (e4) {\(\braket{H}\)};
\vertex at ($(e4) + (-1cm,0)$) (e3) {\(\braket{H}\)};
\vertex at ($(e4) + (1cm,0)$) (e5) {\(\braket{H}\)};

\vertex at ($(e4) + (0,-1cm)$) (d3);

\diagram* {
%Class 1
(a1) -- [fermion] (b1),
(a2) -- [fermion] (b1),
(b1) -- [scalar, edge label=\(\Delta\)] (d1),
(d1) -- [scalar] (e1),
(d1) -- [scalar] (e2),

%Class 2
(a3) -- [fermion] (b2),
(b2) -- [fermion, edge label'=\(\Sigma\)] (b3),
(a4) -- [fermion] (b4),
(b3) -- [anti fermion, edge label'=\(\Sigma\), insertion=0] (b4),
(b2) -- [scalar] (d2),
(b4) -- [scalar, edge label'=\(\Phi_{3/2}\)] (d3),
(e3) -- [scalar,] (d3),
(d3) -- [scalar] (e4),
(e5) -- [scalar] (d3)
};
\end{feynman}
\end{tikzpicture}\\
\vspace{3mm}
\begin{tikzpicture}
\node[] at (2cm,-0.75cm) {$\mathbf{(c)}$};
\node[] at (8cm,-0.75cm) {$\mathbf{(d)}$};
\begin{feynman}
%Class 3
\vertex (a1) {\(\nu_{\alpha L}\)};
\vertex[right=4cm of a1] (a3) {\(\nu_{\beta L}\)};
\vertex at ($(a1)!0.5!(a3)$) (a2) {\(\braket{H}\)};
\vertex at ($(a1)!0.2!(a3) + (0,1.0cm)$) (b1);
\vertex at ($(a1)!0.8!(a3) + (0,1.0cm)$) (b3);
\vertex at ($(b1)!0.5!(b3)$) (b2);

\vertex [above=4cm of a2] (e2) {\(\braket{H}\)};
\vertex at ($(e2) + (-1cm,0)$) (e1) {\(\braket{H}\)};
\vertex at ($(e2) + (1cm,0)$) (e3) {\(\braket{H}\)};

\vertex at ($(e2) + (0,-1.0cm)$) (d1);
\vertex at ($(b2)!0.35!(e2)$) (c1);

%Class 4
\vertex[right=1cm of a3] (a4) {\(\nu_{\alpha L}\)};
\vertex[right=6cm of a4] (a7) {\(\nu_{\beta L}\)};
\vertex at ($(a4)!0.2!(a7) + (0,1.0cm)$) (b4);
\vertex at ($(a4)!0.8!(a7) + (0,1.0cm)$) (b8);
\vertex at ($(b4)!0.25!(b8)$) (b5);
\vertex at ($(b4)!0.5!(b8)$) (b6);
\vertex at ($(b4)!0.75!(b8)$) (b7);
\vertex at ($(b5) + (0,-1.0cm)$) (a5) {\(\braket{H}\)};
\vertex at ($(b7) + (0,-1.0cm)$) (a6) {\(\braket{H}\)};

\vertex [above=2cm of b6] (c2);

\diagram* {
%Class 3
(a1) -- [fermion] (b1),
(b1) -- [fermion, edge label'=\(u_{R}\)] (b2),
(a3) -- [fermion] (b3),
(b2) -- [fermion, edge label'=\(u_{L}\)] (b3),
(b2) -- [scalar] (a2),
(b1) -- [scalar, edge label=\(R_{2}\), quarter left] (c1),
(b3) -- [scalar, edge label'=\(S_{3}\), quarter right] (c1),
(c1) -- [scalar, edge label=\(\Phi_{3/2}\)] (d1),
(e1) -- [scalar] (d1),
(e2) -- [scalar] (d1),
(e3) -- [scalar] (d1),

%Class 4
(a4) -- [fermion] (b4) -- [anti fermion, edge label'=\(\ell_{L}\)] (b5) -- [anti fermion, edge label'=\(\ell_{R}\)] (b6) -- [fermion, edge label'=\(\ell_{R}\)] (b7) -- [fermion, edge label'=\(\ell_{L}\)] (b8) -- [anti fermion] (a7),
(b4) -- [anti charged scalar, edge label=\(\eta^{+}\), quarter left] (c2),
(b8) -- [anti charged scalar, edge label'=\(\eta^{+}\), quarter right] (c2),
(b6) -- [charged scalar, edge label=\(\kappa^{++}\)] (c2),
(b5) -- [scalar] (a5),
(b7) -- [scalar] (a6),
};
\end{feynman}
\end{tikzpicture}
\caption{Overview of neutrino mass-generating processes in (a)~type-II seesaw model, (b)~BNT model, (c)~leptoquark model, and (d) Zee-Babu model.}
\label{Fig:NeutrinoMassModels_FeynmanDiagrams}
\end{figure}

For instance, introducing an $SU(2)_{L}$ triplet scalar field gives rise to the dimension-five operator~$\mathcal{O}_{1}$ at the tree level, as depicted in Fig.~\ref{Fig:NeutrinoMassModels_FeynmanDiagrams}a.
%%%%%%%%%%%%%%%%%%%%%%%%%%%%%%%%%%%%%%%%%%%%%%%%%%%%%%%
%%%%%%%%%%%%%%%%%%%%%%%%%%%%%%%%%%%%%%%%%%%%%%%%%%%%%%%
%%%%%%%%%%%%%%%%%%%%%%%%%%%%%%%%%%%%%%%%%%%%%%%%%%%%%%%%%%%%%%%%%%%%%%%%%%%%%%%%%%%%%%%%%%%%%%%%%%%%%%%%%%%%%%%%%%%%%%%%%%%%%%%%%%%%%%%%%%%%%%%%
Alternatively, considering a higher-dimensional operator of dimension seven, such as $\mathcal{O}_2=LLHH(H^{\dagger}H)/\Lambda^3$, allows the generation of neutrino masses and mixings at the tree level. This can be realised by an $SU(2)_{L}$ quadruplet scalar, including a doubly charged scalar, as shown in Fig.~\ref{Fig:NeutrinoMassModels_FeynmanDiagrams}b.
In certain alternative neutrino mass models, lepton number-violating operators may be absent at the tree level, due to inherent symmetries or the particle content involved. In such radiative mass models, small Majorana masses can be induced at the loop level and the interplay of chirality and loop suppression allows for potentially lower new-physics scales, as demonstrated in Fig.~\ref{Fig:NeutrinoMassModels_FeynmanDiagrams}c, where the neutrino mass arises at the one-loop level by augmenting the SM scalar sector by an $SU(2)_{L}$~quadruplet scalar.
Additionally, we examine the Zee-Babu model, depicted in Fig.~\ref{Fig:NeutrinoMassModels_FeynmanDiagrams}d, where neutrino masses and mixings are generated at the two-loop level, requiring the presence of an $SU(2)_{L}$ singlet doubly-charged scalar, denoted as $\kappa^{++}$.

All these neutrino mass models share a common feature: the presence of doubly charged scalars. A comprehensive list includes type-II seesaw model~\cite{Magg:1980ut, Schechter:1980gr, Lazarides:1980nt, Mohapatra:1980yp}, left-right symmetric model~(LRSM)~\cite{Pati:1974yy, Mohapatra:1974gc, Senjanovic:1975rk, Mohapatra:1979ia}, radiative neutrino mass model~\cite{Zee:1985id, Babu:1988ki}, little Higgs model~\cite{Arkani-Hamed:2002iiv}, $d=7$ neutrino mass models~\cite{Babu:2009aq, Bonnet:2009ej}, $331$~model~\cite{Pisano:1992bxx}, and the Georgi-Machacek~model~\cite{Georgi:1985nv}. In most instances, they manifest themselves either as $SU(2)_{L}$ singlets with hypercharge $Y=2$, as seen in the Zee-Babu model for neutrino masses, or as members of a triplet of $SU(2)_{L}$, as in the type-II seesaw models. In the LRSM, they can exist as either $SU(2)_{L}$-singlet or triplet. However, they may also arise from a doublet, quadruplet, or quintuplet of $SU(2)_{L}$~\cite{Babu:2009aq, Bonnet:2009ej, Babu:2019mfe, Babu:2020hun, Bhattacharya:2016qsg}.

The fact that BSM is required both for the generation of neutrino mass and for DM genesis, a possible connection may very well exist. In the type-I seesaw scenario, for instance, when at least two SM singlet fermions are added to explain neutrino masses, adding another fermion as sterile neutrino DM serves both purposes~\cite{Dodelson:1993je, Shi:1998km, Drewes:2016upu}. However, this sterile neutrino dark matter is isolated from the other, neutrino mass-generating SM singlet fermions. Consequently, it does not exhibit a direct one-to-one correlation to the neutrino masses, and no flavor-specific dark matter signature directly dependent on neutrino oscillation data should be expected. In scenarios where the smallness of neutrino masses is aided by a loop suppression, a BSM symmetry or minuscule charge of particles within the loop that enforces this suppression might also play a role in stabilizing DM~\cite{Tao:1996vb, Ma:2006km, Jana:2024iig}. In such scenarios, one can obtain neutrinophilic dark matter, where dark matter primarily annihilates into SM neutrinos \cite{Herms:2023cyy}. However, establishing any flavor-specific dark matter indirect signatures directly correlated with neutrino oscillation data will be challenging. The reason is that, although both dark matter and neutrino masses can be accommodated within a single framework, dark matter annihilation into neutrinos does not necessarily involve the neutrino mass matrix, unlike what is depicted in Fig.~\ref{Fig:SchematicAnnihilation}.

On the other hand, BSM particles responsible for generating neutrino mass could potentially connect to DM, offering intriguing new physics that establishes flavor-specific DM indirect signatures directly correlated with neutrino oscillation data. Our work investigates the role of such scalars introduced for neutrino mass generation in DM phenomenology. We focus on a minimal scalar DM candidate stabilized by a discrete $\mathbb{Z}_2$ symmetry. Generally, the DMDD constraints restrict this scenario to DM masses in the vicinity of Higgs resonance $m_{\mathrm{DM}} \sim m_{h}/2$ or in a narrow range around $3 \TeV$~\cite{Cirelli:2024ssz}. The scalars responsible for neutrino mass generation, however, couple to the scalar DM (we call it `scalar-neutrino-portal') and provide additional annihilation channels that are not directly connected to DMDD, thus reducing the constraints significantly. On the other hand, the doubly charged scalars could predominantly couple to leptons and provide flavor specific signals both in collider and indirect search experiments, shedding light on neutrino properties like mass ordering and Majorana phases.

%%%%%%%%%%%%%%%%%%%%%%%%%%%%%%%%%%%%%%%%%%%%%%%%%%%%%%%%%%%%%%%%%%%%%%%%%%
\section{Scalar-neutrino-portal in the type-II seesaw model}
\label{Sec:DMinTypeII}
Our focus here will be to elaborate on the \textit{scalar-neutrino-portal} in seesaw models in which neutrino masses are generated at tree-level. As a specific case study, we scrutinize the type-II seesaw framework, wherein the scalar sector is expanded to include an $SU(2)_{L}$ scalar triplet~$\Delta$ with a hypercharge $Y=+1$ alongside a complex scalar singlet~$S$, which undergoes $S \rightarrow -S$ under a  discrete~$\mathbb{Z}_{2}$ symmetry and accounts for DM.\footnote{A real scalar works equally, complex scalar is just a choice with an additional degree of freedom.} The SM fermionic and gauge sectors remain unaltered. The general renormalizable scalar potential reads
\begin{align}
    V_{\mathrm{scalar}} &= -\mu_{H}^{2} H^{\dagger}H + \lambda_{H} \lv H^{\dagger}H \rv^{2} + \mu_{S}^{2} S^{\dagger}S + \lambda_{S} \left( S^{\dagger}S \right)^{2} + \mu_{\Delta}^{2} \mathrm{Tr} \left( \Delta^{\dagger}\Delta \right) 
     \nonumber \\
    &\hspace{1mm}+ \lambda_{S\Delta} S^{\dagger}S \mathrm{Tr} \left( \Delta^{\dagger} \Delta \right) + \lambda_{SH}S^{\dagger}SH^{\dagger}H + \lambda_{H\Delta} \left[ H^{\dagger}H \mathrm{Tr} \left( \Delta^{\dagger} \Delta \right) + \kappa_{H\Delta} \lrvert{H^{\dagger}\Delta}^{2} \right] \nonumber \\
    &\hspace{1mm}+ \mu_{H\Delta} \left( i H^{\top} \sigma_{2}\Delta^{\dagger} H + \mathrm{h.c.} \right)+ \lambda_{\Delta} \left[ \mathrm{Tr}^{2} \left( \Delta^{\dagger} \Delta \right) + \kappa_{\Delta} \mathrm{Tr} \left( \Delta^{\dagger} \Delta \Delta^{\dagger} \Delta \right) \right] \, ,
    \label{Eq:PotentialBSM}
\end{align}
where the SM Higgs field $H$ and scalar triplet field $\Delta$ can be parameterized as
\begin{align}
    H = \frac{1}{\sqrt{2}} \begin{pmatrix} \sqrt{2}G^{+} \\ v_{H} + h + iG^{0} \end{pmatrix} \quad , \quad \Delta = \frac{1}{\sqrt{2}} \begin{pmatrix}
    \Delta^{+} & \sqrt{2}\Delta^{++} \\
    v_{\Delta} + \Delta_{R}^{0} + i\Delta_{I}^{0} & -\Delta^{+}
    \end{pmatrix}
    \label{Eq:MultipletDefinitions}
\end{align}
with the would-be Goldstones~$G^{\pm}$ and $G^{0}$. After electroweak symmetry breaking, the neutral components of the scalar fields~$H$ and $\Delta$ acquire the vacuum expectation values~(vevs) $v_{H}$ and $v_{\Delta}$, respectively, with $v_{H}^{2} + 2v_{\Delta}^{2} = v^{2} \approx (246\GeV)^{2}$. The value of $v_{\Delta}$ can be derived from the minimization condition as
\begin{align}
    v_{\Delta} \approx \frac{\mu_{H\Delta}v_{H}^{2}}{\sqrt{2}\mu_{\Delta}^{2}} \ .
\end{align}
The constraints on the triplet vev are stringent due to its impact on the electroweak $\rho$ parameter, defined as $\rho \equiv m_{W}^{2} / \left( m_{Z}^{2} \cos^{2} \theta_{W} \right)$. In this scenario, the $\rho$ parameter is influenced by $v_{\Delta}$ through $\rho = (1+\frac{2 v_\Delta^2}{v_H^2})/(1+\frac{4 v_\Delta^2}{v_H^2})$. Current precision measurements of electroweak phenomena yield a global fit value for $\rho$ of $1.00038(20)$~\cite{ParticleDataGroup:2022pth}, imposing the upper limit~$v_{\Delta} \lesssim 2\GeV$.

The Yukawa interaction associated with the scalar triplet $\Delta$, which governs neutrino mass, can be written as
\begin{align}
    \mathcal{L}_{\mathrm{Yuk,BSM}} = -i \left( y_{L\Delta} \right)_{\alpha\beta} \overline{L_{\alpha}^{c}} \sigma_{2} \Delta L_{\beta} + \mathrm{h.c.} \, , \label{Eq:YukawaBSM}
\end{align}
with the Hermitian $3\times 3$ Yukawa matrix~$y_{L\Delta}$, the Pauli matrix~$\sigma_{2}$, and the left-handed lepton doublet~$L_{\alpha}$, where $\alpha$ denotes the flavor index $\alpha=e,\mu,\tau$. When the neutral component of the triplet scalar obtains a vev, i.e.~$\braket{\Delta^{0}} \equiv v_{\Delta}/\sqrt{2} > 0$, it generates a Majorana mass for the neutrinos through the Yukawa term as
\begin{align}
    \left( m_{\nu} \right)_{ij} = \sqrt{2} v_{\Delta} \left( y_{L\Delta} \right)_{ij} = \left[ V_{\mathrm{PMNS}}^{\mathrm{M}} \, \mathrm{diag}\left(m_{1}, m_{2}, m_{3} \right)\left( V_{\mathrm{PMNS}}^{\mathrm{M}} \right)^{T} \right]_{ij} \, ,
    \label{Eq:mass}
\end{align}
where $V_{\mathrm{PMNS}}^{\mathrm{M}}$ represents the Pontecorvo-Maki-Nakagawa-Sakata (PMNS) matrix
\begin{align}
    V_{\mathrm{PMNS}}^{\mathrm{M}} = \begin{pmatrix}
    c_{12}c_{13} & s_{12}c_{13} & s_{13}e^{-i\delta_{\mathrm{CP}}} \\
    -s_{12}c_{23} - c_{12}s_{23}s_{13}e^{i\delta_{\mathrm{CP}}} & c_{12}c_{23} - s_{12}s_{23}s_{13}e^{i\delta_{\mathrm{CP}}} & s_{23}c_{13} \\
    s_{12}s_{23} - c_{12}c_{23}s_{13}e^{i\delta_{\mathrm{CP}}} & -c_{12}s_{23} - s_{12}c_{23}s_{13}e^{i\delta_{\mathrm{CP}}} & c_{23}c_{13}
    \end{pmatrix} \times \mathcal{D}^{\mathrm{M}}
    \label{Eq:PMNSmatrixDirac}
\end{align}
with the diagonal matrix~$\mathcal{D}^{\mathrm{M}} \equiv \mathrm{diag}\left( 1 , e^{i\varphi_{1}/2} , e^{i\varphi_{2}/2} \right)$, $s_{ij}\equiv \sin\theta_{ij}$, $c_{ij}\equiv \cos\theta_{ij}$, the CP-violating Dirac phase $\delta_{\mathrm{CP}}$, and two Majorana phases $\varphi_{1,2}$. Following Eq.~(\ref{Eq:mass}), it becomes apparent that the Yukawa matrix configuration can be fully determined by the neutrino mass matrix, thus allowing for its reformulation as
\begin{align}
    \label{Eq:SSIIyukawas}
    \left( y_{L\Delta} \right)_{ij} = \frac{1}{\sqrt{2} v_{\Delta}} \left[ V_{\mathrm{PMNS}}^{\mathrm{M}} \, \mathrm{diag}\left(m_{1}, m_{2}, m_{3} \right)\left( V_{\mathrm{PMNS}}^{\mathrm{M}} \right)^{T} \right]_{ij} \, .
\end{align}
The elements of the Yukawa matrix~$y_{L\Delta}$ corresponding to the three distinct flavors can consequently be expressed as \cite{Akeroyd:2007zv}
\begin{align}
    \left( y_{L\Delta} \right)_{ee} &= \frac{1}{\sqrt{2}v_{\Delta}} \left(m_{1}c_{12}^{2} c_{13}^{2} + m_{2} c_{13}^{2}s_{12}^{2} e^{i\varphi_{1}} + m_{3} s_{13}^{2} e^{i\left(\varphi_{2}-2\delta_{\mathrm{CP}}\right)}\right)\,, \nonumber \\
    \nonumber \\
    \left( y_{L\Delta} \right)_{e\mu} &= \frac{1}{\sqrt{2}v_{\Delta}} \left[-m_{1}c_{12} c_{13} \left( s_{12}c_{23} + c_{12}s_{13}s_{23}e^{i\delta_{\mathrm{CP}}} \right) \right.\nonumber\\
    &\hspace{5mm} \left.+ m_{2} c_{13}s_{12} e^{i\varphi_{1}} \left( c_{12}c_{23} - s_{12}s_{13}s_{23}e^{i\delta_{\mathrm{CP}}} \right) + m_{3} s_{13}s_{23}c_{13}e^{i\left(\varphi_{2}-\delta_{\mathrm{CP}}\right)} \right]\,, \nonumber \\
    \nonumber \\
    \left( y_{L\Delta} \right)_{e\tau} &= \frac{1}{\sqrt{2}v_{\Delta}} \left[m_{1}c_{12} c_{13} \left( s_{12}s_{23} - c_{12}c_{23}s_{13}e^{i\delta_{\mathrm{CP}}} \right) \right.\nonumber\\
    &\hspace{5mm} \left.- m_{2} c_{13}s_{12} e^{i\varphi_{1}} \left( c_{12}s_{23} + s_{12}s_{13}c_{23}e^{i\delta_{\mathrm{CP}}} \right) + m_{3} s_{13}c_{23}c_{13}e^{i\left(\varphi_{2}-\delta_{\mathrm{CP}}\right)} \right]\,, \nonumber \\
    \nonumber \\
    \left( y_{L\Delta} \right)_{\mu\mu} &= \frac{1}{\sqrt{2}v_{\Delta}} \left[m_{1} \left( s_{12}c_{23} + c_{12}s_{23}s_{13}e^{i\delta_{\mathrm{CP}}} \right)^{2} \right.\nonumber\\
    &\hspace{5mm} \left.+ m_{2}e^{i\varphi_{1}} \left( c_{12}c_{23} - s_{12}s_{13}s_{23}e^{i\delta_{\mathrm{CP}}} \right)^{2} + m_{3} c_{13}^{2}s_{23}^{2}e^{i\varphi_{2}} \right]\,, \nonumber \\
    \nonumber \\
    \left( y_{L\Delta} \right)_{\mu\tau} &= \frac{1}{\sqrt{2}v_{\Delta}} \left[m_{1} \left( c_{12}c_{23}s_{13}e^{i\delta_{\mathrm{CP}}} - s_{12}s_{23} \right)\left( s_{12}c_{23} + c_{12}s_{23}s_{13}e^{i\delta_{\mathrm{CP}}} \right) \right.\nonumber\\
    &\hspace{5mm} +m_{2}e^{i\varphi_{1}} \left( c_{12}s_{23} + s_{12}c_{23}s_{13}e^{i\delta_{\mathrm{CP}}} \right)\left( s_{12}s_{23}s_{13}e^{i\delta_{\mathrm{CP}}} - c_{12}c_{23} \right) \nonumber \\
    &\hspace{5mm} \left. + m_{3} c_{13}^{2}c_{23}s_{23}e^{i\varphi_{2}} \right]\,, \nonumber \\
    \nonumber \\
    \left( y_{L\Delta} \right)_{\tau\tau} &= \frac{1}{\sqrt{2}v_{\Delta}} \left[m_{1} \left( s_{12}s_{23} - c_{12}c_{23}s_{13}e^{i\delta_{\mathrm{CP}}} \right)^{2} \right.\nonumber\\
    &\hspace{5mm} \left.+ m_{2}e^{i\varphi_{1}} \left( c_{12}s_{23} + s_{12}s_{13}c_{23}e^{i\delta_{\mathrm{CP}}} \right)^{2} + m_{3} c_{13}^{2}c_{23}^{2}e^{i\varphi_{2}} \right] \ .
    \label{Eq:YukawaCouplings}
\end{align}

The Yukawa matrix $y_{L\Delta}$ can be deduced from nine parameters governing neutrino oscillations: three mixing angles~$\theta_{12}, \theta_{23}, \theta_{13}$, two mass differences~$\Delta m_{21}^{2}, \Delta m_{3\ell}^{2}$, the lightest neutrino mass~$m_{0} \equiv m_{1(3)}$ for normal hierarchy~(NH) and inverted hierarchy~(IH) respectively, and three CP phases~$\delta_{\mathrm{CP}}, \varphi_{1}, \varphi_{2}$. As presented in Tab.~\ref{Tab:NeutrinoOscillationData}, five neutrino oscillation observables -- namely, the three mixing angles and two mass differences -- have been measured with a precision of a few percent. Additionally, the Dirac CP phase has been the subject of ongoing investigation, notwithstanding some disparities between the observed $\delta_{\mathrm{CP}}$ values from the T2K and NO$\nu$A experiments. Nevertheless, with more data and forthcoming experiments, the measurement of~$\delta_{\mathrm{CP}}$ will achieve heightened precision. This leaves us with three remaining neutrino oscillation parameters: the absolute neutrino mass scale~$m_{0}$ and the two Majorana phases. Neutrino oscillation experiments lack sensitivity to these parameters. However, as subsequent sections shall demonstrate, notable complementarity exists in the indirect detection signals of DM, collider experiments, and flavor-violating observables. Each of these observables exhibits sensitivity to different unknowns and we can glean supplementary insights into neutrino oscillation phenomena through this complementarity. It is important to highlight that, at present, there is no conclusive evidence directly connecting dark matter to neutrinos. However, if both dark matter and neutrinos arise from a unified theoretical framework, dark matter signatures could be correlated with neutrino oscillation observables.

Before analyzing interconnections with DM, it is important to emphasize that the SM predicts the absence of lepton flavor-violating processes. However, an inevitable expectation of lepton flavor violation arises as neutrino mass models emerge to explain oscillation data by generating neutrino masses and mixings. Various neutrino mass models yield different predictions of lepton flavor-violating observables.  Notably, within the type-II seesaw mechanism, thorough analyses of these observables have been conducted~\cite{BhupalDev:2018tox,Primulando:2019evb,Fridell:2023gjx}. Here we present a brief summary of these lepton flavor-violating observables and the constraints placed on the Yukawa matrix $y_{L\Delta}$ since the Yukawa interactions induce lepton flavor-violating phenomena, such as $\ell_a^{-} \rightarrow \ell_b^{+} \ell_c^{-} \ell_d^{-}$ and $\ell_a^{-} \rightarrow \ell_b^{-} \gamma$, which are tightly constrained~(cf. Tab.~\ref{Tab:TypeII_LeptonflavorViolationTreeLevelLimits}).
\begin{table}[b!]
    \centering
    \begin{tabular}{cc}
        \toprule
        LFV observable & Constraints \\
        \midrule
        $\mathrm{BR} \left( \mu^{-} \rightarrow e^{+} e^{-} e^{-} \right) < 1.0\times 10^{-12}$ & $\left\vert \left( m_{\nu} \right)_{e\mu} \left( m_{\nu} \right)_{ee} \right\vert/\left( m_{\Delta^{++}}^{2} v_{\Delta}^{2} \right) < \left(145\TeV \right)^{-2}$ \\
        $\mathrm{BR} \left( \tau^{-} \rightarrow e^{+} e^{-} e^{-} \right) < 2.7\times 10^{-8}$ & $\left\vert \left( m_{\nu} \right)_{e\tau} \left( m_{\nu} \right)_{ee} \right\vert/\left( m_{\Delta^{++}}^{2} v_{\Delta}^{2} \right) < \left(7.4\TeV \right)^{-2}$ \\
        $\mathrm{BR} \left( \tau^{-} \rightarrow e^{+} e^{-} \mu^{-} \right) < 1.8\times 10^{-8}$ & $\left\vert \left( m_{\nu} \right)_{e\tau} \left( m_{\nu} \right)_{e\mu} \right\vert/\left( m_{\Delta^{++}}^{2} v_{\Delta}^{2} \right) < \left(9.8\TeV \right)^{-2}$ \\
        $\mathrm{BR} \left( \tau^{-} \rightarrow e^{+} \mu^{-} \mu^{-} \right) < 1.7\times 10^{-12}$ & $\left\vert \left( m_{\nu} \right)_{e\tau} \left( m_{\nu} \right)_{\mu\mu} \right\vert/\left( m_{\Delta^{++}}^{2} v_{\Delta}^{2} \right) < \left(8.3\TeV \right)^{-2}$ \\
        $\mathrm{BR} \left( \tau^{-} \rightarrow \mu^{+} e^{-} e^{-} \right) < 1.5\times 10^{-8}$ & $\left\vert \left( m_{\nu} \right)_{\mu\tau} \left( m_{\nu} \right)_{ee} \right\vert/\left( m_{\Delta^{++}}^{2} v_{\Delta}^{2} \right) < \left(8.6\TeV \right)^{-2}$ \\
        $\mathrm{BR} \left( \tau^{-} \rightarrow \mu^{+} \mu^{-} e^{-} \right) < 2.7\times 10^{-8}$ & $\left\vert \left( m_{\nu} \right)_{\mu\tau} \left( m_{\nu} \right)_{e\mu} \right\vert/\left( m_{\Delta^{++}}^{2} v_{\Delta}^{2} \right) < \left(8.8\TeV \right)^{-2}$ \\
        $\mathrm{BR} \left( \tau^{-} \rightarrow \mu^{+} \mu^{-} \mu^{-} \right) < 2.1\times 10^{-8}$ & $\left\vert \left( m_{\nu} \right)_{\mu\tau} \left( m_{\nu} \right)_{\mu\mu} \right\vert/\left( m_{\Delta^{++}}^{2} v_{\Delta}^{2} \right) < \left(7.9\TeV \right)^{-2}$  \\
        \midrule
        $\mathrm{BR} \left( \mu \rightarrow e\gamma \right) < 4.2 \times 10^{-13}$ & $\left( 8 + r \right) \left\vert m_{\nu}^{\dagger} m_{\nu} \right\vert_{e\mu} / \left( m_{\Delta^{++}}^{2} v_{\Delta}^{2} \right) < \left( 15.3\TeV \right)^{-2}$ \\
        $\mathrm{BR} \left( \tau \rightarrow e\gamma \right) < 3.3 \times 10^{-8}$ & $\left( 8 + r \right) \left\vert m_{\nu}^{\dagger} m_{\nu} \right\vert_{e\tau} / \left( m_{\Delta^{++}}^{2} v_{\Delta}^{2} \right) < \left( 0.6\TeV \right)^{-2}$ \\
        $\mathrm{BR} \left( \tau \rightarrow \mu\gamma \right) < 4.4 \times 10^{-8}$ & $\left( 8 + r \right) \left\vert m_{\nu}^{\dagger} m_{\nu} \right\vert_{\mu\tau} / \left( m_{\Delta^{++}}^{2} v_{\Delta}^{2} \right) < \left( 0.56\TeV \right)^{-2}$ \\
    \hline
  \end{tabular}
  \caption{Constraints from different lepton flavor-violating~(LFV) processes. The experimental limits at $90\%$ C.L. are taken from Refs.~\cite{SINDRUM:1987nra, Hayasaka:2010np, BaBar:2009hkt, MEG:2016leq, ParticleDataGroup:2018ovx} and the constraints from Ref.~\cite{Primulando:2019evb}. Here we use~$r\equiv m_{\Delta^{++}}^{2}/m_{\Delta^{+}}^{2}$.}
  \label{Tab:TypeII_LeptonflavorViolationTreeLevelLimits}
\end{table}

\begin{figure}[b!]
    \centering
    \includegraphics[width=0.49\textwidth]{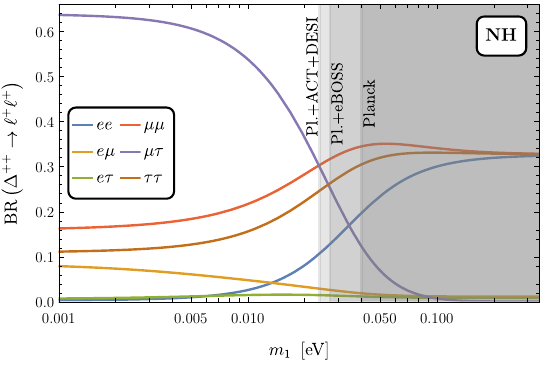}
    \includegraphics[width=0.49\textwidth]{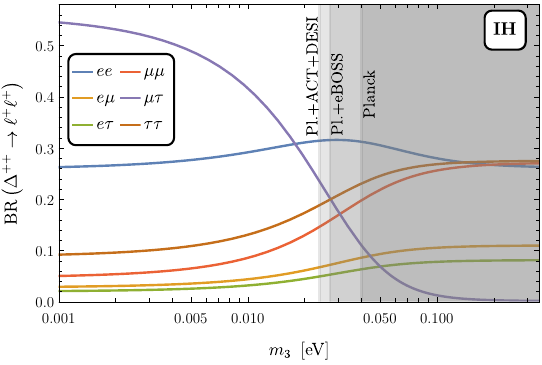}\\
    \includegraphics[width=0.49\textwidth]{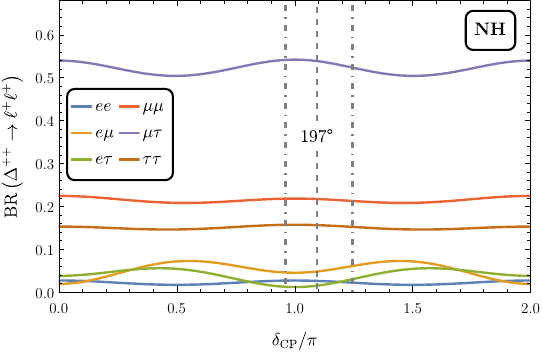}
    \includegraphics[width=0.49\textwidth]{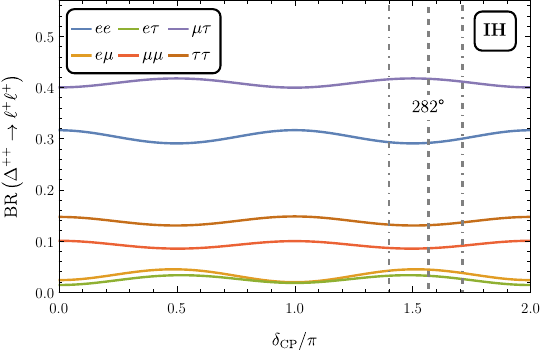}\\
    \includegraphics[width=0.49\textwidth]{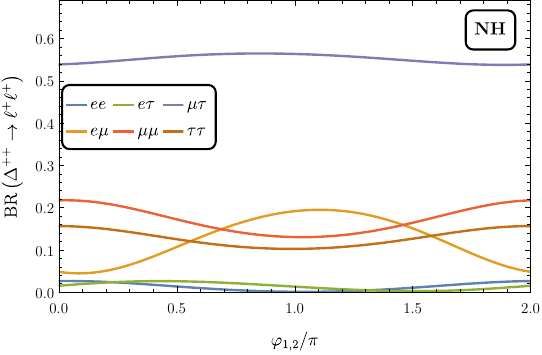}
    \includegraphics[width=0.49\textwidth]{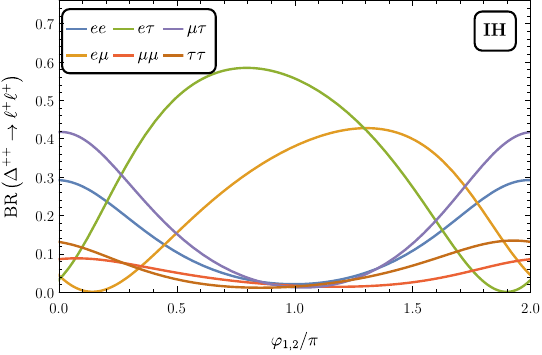}
    \caption{Dependence of the branching ratios on free neutrino oscillation parameters. The BSM vev is set to~$v_{\Delta}=10^{-5}\GeV$, and we choose $\delta_{\mathrm{CP}} = 197^{\circ} \left( 282^{\circ} \right)$ for NH~(IH) as the central values of the CP phase (cf. Tab.~\ref{Tab:NeutrinoOscillationData}). The neutrino mass scale is set to $m_{0}=0.01\eV$ and the Majorana phases to $\varphi_{1,2} = 0$ if not stated otherwise. Note the asymmetry with respect to the Majorana phases due to $\delta_{\mathrm{CP}} \, \mathrm{mod} \, \pi \neq 0$. The mass of the lightest neutrino is constrained by Planck~\cite{Planck:2018vyg}, extended Baryon Oscillation Spectroscopy Survey (eBOSS)~\cite{Brieden:2022lsd}, Atacama Cosmology Telescope (ACT)~\cite{ACT:2023kun}, and Dark Energy Spectroscopic Instrument (DESI)~\cite{DESI:2024mwx}.}
    \label{Fig:TypeII_Delta_BR}
\end{figure}

In DM indirect detection, as we shall elaborate, the interaction between the scalar mediator and different lepton flavors determines the spectrum of cosmic ray positrons expected from DM annihilation in our Galaxy. This link between DM annihilation and SM leptons through generic, doubly or singly charged scalars connects neutrino oscillation observables with flavored DM signatures, as shown in Fig.~\ref{Fig:SchematicAnnihilation}. An essential aspect lies in scrutinizing the decay processes of singly and doubly charged scalars into leptons, which are dictated by the neutrino mass matrix.
Typically, the fields of the triplet scalar~$\Delta$ may feature a mass splitting, giving rise to the potential for two hierarchical configurations:
\begin{align}
\mathrm{Case \ A} \ &: \ m_{\Delta^0} \geq m_{\Delta^{+}} \geq m_{\Delta^{++}} \,,\\
\mathrm{Case \ B} \ &: \ m_{\Delta^{++}}>m_{\Delta^{+}}>m_{\Delta^0} \, .
\end{align}
For simplicity, we consider the scenario where $\Delta^{++}$ is the lightest of the fields from the triplet. Now, the doubly charged scalar decay modes depend on the vev~$v_{\Delta}$ and mass splitting between the fields from the triplet, and this can be nicely presented in decay phase diagrams as shown in Ref.~\cite{Melfo:2011nx}. Since we are concentrating on leptonic decay modes of the doubly charged scalar, we set $v_{\Delta} \ll 10^{-4}\GeV$ such that the leptonic branching fraction is $100\%$ and we set the mass splitting very small such that cascade decays do not open up. The decay rate of the doubly charged scalar~$\Delta^{++}$ into a pair of same-sign charged leptons can be expressed as~\cite{Akeroyd:2007zv,FileviezPerez:2008jbu, Melfo:2011nx, Mandal:2022zmy}
\begin{align}
    \Gamma \left( \Delta^{++} \rightarrow \ell_{i}^{+} \ell_{j}^{+} \right) = \frac{m_{\Delta^{++}}}{4\pi \left( 1+ \delta_{ij} \right)} \vert \left( y_{L\Delta} \right)_{ij} \vert^{2} \,, 
\end{align}
and, consequently, the branching ratio to various flavor leptons is given by
\begin{align}
    \mathrm{BR} \left( \Delta^{++} \rightarrow \ell_{i}^{+} \ell_{j}^{+} \right) = \frac{\vert \left( y_{L\Delta} \right)_{ij} \vert^{2}}{1+ \delta_{ij}} / \sum_{a\leq b} \frac{\lv \left( y_{L\Delta} \right)_{ab} \rv^{2}}{1+ \delta_{ab}} \, .
\end{align}
In Fig.~\ref{Fig:TypeII_Delta_BR}, we illustrate how the branching ratios vary with different neutrino oscillation parameters.
By fixing the values of three mixing angles ($\theta_{12}$, $\theta_{23}$, $\theta_{13}$) and two mass differences ($\Delta m_{21}^{2}$, $\Delta m_{3\ell}^{2}$) to their respective central values from Tab.~\ref{Tab:NeutrinoOscillationData}, it becomes evident that the decay of doubly charged scalar $\Delta^{++}$ into specific flavors of leptons is influenced by the three CP phases ($\delta_{\mathrm{CP}}$, $\varphi_{1}$, $\varphi_{2}$) and the lightest neutrino mass. Furthermore, the expectations for the decay of the doubly charged scalar into different lepton flavors change according to the neutrino mass ordering. Consequently, we can derive additional insights into neutrino oscillation phenomena.

We have outlined how neutrino observables influence the interactions between prospective scalar mediators and various lepton flavors within the type-II seesaw framework, which will have implications for DM indirect detection.
This discussion extends to other tree-level seesaw models, in particular to those involving generic $SU(2)_{L}$ multiplets. In such cases, the emergence of a vev of the neutral multiplet component results in Majorana neutrino mass terms via Yukawa interactions.  The Yukawa texture is then directly correlated to neutrino oscillation observables. For example, in another tree-level $d=7$ neutrino mass model, known as the BNT model\footnote{See Refs.~\cite{Babu:2009aq, Ghosh:2017jbw, Ghosh:2018drw, Bhattacharya:2016qsg} for more details.}, the neutral component of an $SU(2)_{L}$ quadruplet develops a vev (cf. Fig.~\ref{Fig:NeutrinoMassModels_FeynmanDiagrams}b). This quadruplet comprises triply, doubly, and singly charged scalars besides the neutral one whose Yukawa interactions generate the neutrino flavor structure as we know from neutrino oscillation data. Similar to the type-II seesaw framework, Yukawa couplings for such quadruplet scalars can also be written in terms of neutrino oscillation observables (cf.~Eq.~(\ref{Eq:YukawaCouplings})), affecting the charged scalar branching fractions to different lepton flavors impacting both cosmic ray positron spectra and collider observations.

\section{Scalar-neutrino-portal in radiative neutrino mass models}
\label{Sec:DMinZeeBabu}
In this section, we will explore the implications of new scalar mediators originating from radiative neutrino mass models, where small Majorana masses can emerge at the loop level, and the combination of chirality and loop suppression can allow for potentially lower new physics scales. Generally, these models require only an extension of the scalar sector.
As a case study, we will examine the Zee-Babu model~\cite{Zee:1985id, Babu:1988ki}, where neutrino masses and mixings are generated at the two-loop level. In this model, the particle spectrum remains minimal, extending SM with two $SU(2)_{L}$ singlet scalar fields, $\eta^{+}: (\mathbf{1},\mathbf{1},1)$ and $\kappa^{++}: (\mathbf{1},\mathbf{1},2)$, which are singly and doubly charged, respectively. A complex scalar singlet $S$, stabilized by a $\mathbb{Z}_{2}$ symmetry, accounts for DM as before. The relevant kinematic and scalar potential terms are as follows:
\begin{align}
    \mathcal{L} &\supset \sum_{\phi} \left[ \left( D_{\mu} \phi \right)^{\dagger} \left( D^{\mu} \phi \right) - \mu_{\phi}^{2} \phi^{\dagger} \phi - \lambda_{\phi} \left( \phi^{\dagger} \phi \right)^{2} \right]
    - \left( \mu_{\eta\kappa} \eta^{+}\eta^{+} \kappa^{--} + \mathrm{h.c.} \right) \nonumber \\
    &\hspace{5mm}- \lambda_{SH} S^{\dagger} S \vert H \vert^{2} - \lambda_{H\eta} \eta^{-} \eta^{+} \vert H \vert^{2} - \lambda_{H\kappa} \kappa^{--} \kappa^{++} \vert H \vert^{2} - \lambda_{\eta \kappa} \eta^{-} \eta^{+} \kappa^{--} \kappa^{++} \nonumber \\
    &\hspace{5mm}- \lambda_{S\eta} \eta^{-}\eta^{+} S^{\dagger}S - \lambda_{S\kappa} \kappa^{--}\kappa^{++} S^{\dagger}S\,,
\end{align}
with the sum over $\phi = H, S, \eta^{+}, \kappa^{++}$. 

The Yukawa Lagrangian, featuring $\eta^{+}$ and $\kappa^{++}$, can be expressed as follows:
\begin{align}
    \mathcal{L}_{\mathrm{Yuk,BSM}} \supset  f_{i j} L_i^{a T} C L_j^b \epsilon_{a b} \eta^{+}+g_{i j} \ell_i^T C \ell_j \kappa^{++} + \mathrm{h.c.} \, ,
\end{align}
where $C$ refers to the charge-conjugation matrix, $L$ represents the left-handed lepton doublet, and $\ell$ denotes the right-handed charged lepton singlet. The indices $(ij)$ and $(ab)$ correspond to the generation and $SU(2)$ indices, respectively. The Yukawa coupling $f_{ij}$ is antisymmetric ($f_{ij} = -f_{ji}$) due to the Clebsch factor $\epsilon_{ab}$, whereas $g_{ij}$ is symmetric ($g_{ij} = g_{ji}$). 

Assuming the new scalar fields $\eta^{+}$ and $\kappa^{++}$ possess lepton numbers of two units, specifically $L(\eta^+) = -2$ and $L(\kappa^{++}) = -2$, all terms in the Lagrangian conserve lepton number except for the term $\mu_{\eta\kappa} \eta^{+} \eta^{+} \kappa^{--}$ in the scalar potential. This results in a breaking of lepton number by two units, enabling the generation of Majorana neutrino masses at the two-loop level, as shown in Fig.~\ref{Fig:NeutrinoMassModels_FeynmanDiagrams}d. The neutrino mass matrix is given by~\cite{Nebot:2007bc, Schmidt:2014zoa}
\begin{align}
    m_{\nu} = \frac{8\mu_{\eta\kappa} v^{2}}{ \left( 16 \pi^{2} \right)^{2} m_{\eta}^{2}} \, f \, y \, g^{\dagger} \, y^{T} \, f^{T} \,  \mathcal{I} \left( \frac{m_{\kappa}^{2}}{m_{\eta}^{2}}\right), \label{Eq:NuMassMatrix_ZeeBabu}
\end{align}
where the loop integral function can be expressed as~\cite{AristizabalSierra:2006gb} 
\begin{align}
    \mathcal{I} \left( r \right) = - \int_{0}^{1} \dd z \int_{0}^{1-z} \dd y \frac{1}{z + (r-1)y + y^{2}} \log \frac{y \left( 1-y \right)}{z + ry} \, .
\end{align}
The integral can be approximated in certain limits,
\begin{align}
    \mathcal{I} \left( r \right) \approx \begin{cases} \frac{1}{r} \left( \log^{2} r + \frac{\pi^{2}}{3} - 1 \right) & \mathrm{for} \ r \gg 1\,, \\ \frac{\pi^{2}}{3} & \mathrm{for} \ r \rightarrow 0\,, \end{cases}
\end{align}
where~$r\equiv m_{\kappa}^{2}/m_{\eta}^{2}$.

The neutrino mass matrix $m_{\nu}$ is connected to the PMNS matrix as described in Eq.~(\ref{Eq:PMNSmatrixDirac}), thereby establishing relationships between neutrino oscillation data and the Yukawa couplings as 
\begin{align}
    m_{\nu}^{\mathrm{NH,IH}} = V_{\mathrm{PMNS}}^{\mathrm{D}} \, \mathcal{D}_{\nu}^{\mathrm{NH,IH}} \left( V_{\mathrm{PMNS}}^{\mathrm{D}} \right)^{T} \, ,
\end{align}
with the diagonal matrices~$\mathcal{D}_{\nu}^{\mathrm{NH}} = \mathrm{diag}\left(0, m_{2}e^{i\varphi} , m_{3} \right)$ and $\mathcal{D}_{\nu}^{\mathrm{IH}} = \mathrm{diag}\left(m_{1}, m_{2}e^{i\varphi} , 0 \right)$ for normal and inverted hierarchy, respectively. The skew-symmetric matrix~$f$ can be expressed in terms of the neutrino oscillation data (see Ref.~\cite{Nebot:2007bc}). Since the eigenvector~$\boldsymbol{f}_{\lambda=0} = \left( f_{\mu\tau}, -f_{e\tau}, f_{e\mu} \right)$ of the matrix~$f$, associated with the eigenvalue~$\lambda = 0$, is also an eigenvector of the neutrino mass matrix, one finds
\begin{align}
    \frac{f_{e\tau}}{f_{\mu\tau}} = t_{12} \frac{c_{23}}{c_{13}} + t_{13} s_{23} e^{-i\delta_{\mathrm{CP}}} \quad , \quad \frac{f_{e\mu}}{f_{\mu\tau}} = t_{12} \frac{s_{23}}{c_{13}} - t_{13} c_{23} e^{-i\delta_{\mathrm{CP}}}\,,
    \label{Eq:fMatrixComponentsNH}
\end{align}
 for normal hierarchy, and
\begin{align}
    \frac{f_{e\tau}}{f_{\mu\tau}} = -\frac{s_{23}}{t_{13}} e^{-i\delta_{\mathrm{CP}}} \quad , \quad \frac{f_{e\mu}}{f_{\mu\tau}} = \frac{c_{23}}{t_{13}} e^{-i\delta_{\mathrm{CP}}}
    \label{Eq:fMatrixComponentsIH}\,,
\end{align}
for inverted hierarchy, where again the abbreviations~$s_{ij}$, $c_{ij}$, and $t_{ij}$ denote $\sin\theta_{ij}$, $\cos\theta_{ij}$, and $\tan\theta_{ij}$, respectively. Note that~$f_{e\mu}$ and $f_{e\tau}$ are inversely proportional to $\sin\theta_{13}$ for a fixed~$f_{\mu\tau}$, which has important implications for lepton flavor-violating processes. In order to find the relations for the elements of matrix~$g$ we consider the neutrino mass matrix with the prefactor~$\xi \equiv \mu_{\eta\kappa} v^{2} \mathcal{I} \left( m_{\kappa}^{2} / m_{\eta}^{2}\right) / \left( 32 \pi^{4} m_{\eta}^{2} \right) $ from Eq.~(\ref{Eq:NuMassMatrix_ZeeBabu}).
As three more parameters are to be determined, only three equations must be chosen. Following Ref.~\cite{Nebot:2007bc}, one set of equations can read
\begin{align}
    \left( m_{\nu} \right)_{22} &= \frac{2\xi}{v^{2}} \left( f_{e\mu}^{2} g_{ee}^{*} m_{e}^{2} - 2 f_{e\mu} f_{\mu\tau} g_{e\tau}^{*} m_{e} m_{\tau} + f_{\mu\tau}^{2} g_{\tau\tau}^{*} m_{\tau}^{2} \right)\,, \\
    \left( m_{\nu} \right)_{23} &= \frac{2\xi}{v^{2}} \left( f_{e\mu} f_{e\tau} g_{ee}^{*} m_{e}^{2} + f_{e\mu} f_{\mu\tau} g_{e\mu}^{*} m_{e} m_{\mu} - f_{e\tau} f_{\mu\tau} g_{e\tau}^{*} m_{e} m_{\tau} - f_{\mu\tau}^{2} g_{\mu\tau}^{*} m_{\mu} m_{\tau} \right)\,, \\
    \left( m_{\nu} \right)_{33} &= \frac{2\xi}{v^{2}} \left( f_{e\tau}^{2} g_{ee}^{*} m_{e}^{2} + 2 f_{e\tau} f_{\mu\tau} g_{e\mu}^{*} m_{e} m_{\mu} + f_{\mu\tau}^{2} g_{\mu\mu}^{*} m_{\mu}^{2} \right)\,,
\end{align}
and the entries of the neutrino mass matrix on the left-hand sides are given by
\begin{align}
    \left( m_{\nu}^{\mathrm{NH}} \right)_{22} &= m_{3} c_{13}^{2} s_{23}^{2} + m_{2} e^{i \varphi} \left( c_{12} c_{23} - s_{12} s_{13} s_{23} e^{i \delta_{\mathrm{CP}}} \right)^{2}\,, \\
    \left( m_{\nu}^{\mathrm{NH}} \right)_{23} &= m_{3} c_{13}^{2} s_{23} c_{23} - m_{2} e^{i \varphi} \left( c_{12} s_{23} + s_{12} s_{13} c_{23} e^{i \delta_{\mathrm{CP}}} \right) \left( c_{12} c_{23} - s_{12} s_{13} s_{23} e^{i \delta_{\mathrm{CP}}} \right)\,, \\
    \left( m_{\nu}^{\mathrm{NH}} \right)_{33} &= m_{3} c_{13}^{2} c_{23}^{2} + m_{2} e^{i \varphi} \left( c_{12} s_{23} + s_{12} s_{13} c_{23} e^{i \delta_{\mathrm{CP}}} \right)^{2}\,,
\end{align}
for the normal hierarchy and
\begin{align}
    \left( m_{\nu}^{\mathrm{IH}} \right)_{22} &= m_{1} \left( s_{12} c_{23} + c_{12} s_{13} s_{23} e^{i \delta_{\mathrm{CP}}} \right)^{2} + m_{2} e^{i \varphi} \left( c_{12} c_{23} - s_{12} s_{13} s_{23} e^{i \delta_{\mathrm{CP}}} \right)^{2}\,, \\
    \left( m_{\nu}^{\mathrm{IH}} \right)_{23} &= m_{1} \left( -s_{12} s_{23} + c_{12} s_{13} c_{23} e^{i \delta_{\mathrm{CP}}} \right) \left( s_{12} c_{23} + c_{12} s_{13} s_{23} e^{i \delta_{\mathrm{CP}}} \right) \nonumber \\
    &\hspace{5mm} + m_{2} e^{i \varphi} \left( -c_{12} c_{23} + s_{12} s_{13} s_{23} e^{i \delta_{\mathrm{CP}}} \right) \left( c_{12} s_{23} + s_{12} s_{13} c_{23} e^{i \delta_{\mathrm{CP}}} \right)\,, \\
    \left( m_{\nu}^{\mathrm{IH}} \right)_{33} &= m_{1} \left( s_{12} s_{23} - c_{12} s_{13} c_{23} e^{i \delta_{\mathrm{CP}}} \right)^{2} + m_{2} e^{i \varphi} \left( c_{12} s_{23} + s_{12} s_{13} c_{23} e^{i \delta_{\mathrm{CP}}} \right)^{2}\,,
\end{align}
for the inverted hierarchy. Solving the set of equations leads to
\begin{align}
    g_{\mu\mu}^{*\mathcal{H}} &= \frac{v^{2}}{2f_{\mu\tau}^{2} m_{\mu}^{2} \xi} \left( m_{\nu}^{\mathcal{H}} \right)_{33} - \frac{f_{e\tau} m_{e}}{f_{\mu\tau}^{2} m_{\mu}} \left( f_{e\tau} g_{ee}^{*} \frac{m_{e}}{m_{\mu}} + 2 f_{\mu\tau} g_{e\mu}^{*} \right)\,, \\
    g_{\mu\tau}^{*\mathcal{H}} &= \frac{f_{e\mu} m_{e} \left( f_{e\tau} g_{ee}^{*} m_{e} + f_{\mu\tau} g_{e\mu}^{*} m_{\mu} \right)}{f_{\mu\tau}^{2} m_{\mu} m_{\tau}} - \frac{f_{e\tau} g_{e\tau}^{*} m_{e}}{f_{\mu\tau} m_{\mu}} - \frac{v^{2}}{2f_{\mu\tau}^{2} m_{\mu} m_{\tau}\xi} \left( m_{\nu}^{\mathcal{H}} \right)_{23}\,, \\
    g_{\tau\tau}^{*\mathcal{H}} &= \frac{v^{2}}{2f_{\mu\tau}^{2} m_{\tau}^{2} \xi} \left( m_{\nu}^{\mathcal{H}} \right)_{22} - \frac{f_{e\mu} m_{e}}{f_{\mu\tau}^{2} m_{\tau}} \left( f_{e\mu} g_{ee}^{*} \frac{m_{e}}{m_{\tau}} - 2 f_{\mu\tau} g_{e\tau}^{*} \right)\,,
\end{align}
for both hierarchies~$\mathcal{H}\in\{\mathrm{NH},\mathrm{IH}\}$. These results match the ones in Ref.~\cite{Nebot:2007bc} for small Yukawa couplings~$g_{ei}$. 

The Yukawa matrices involved in the neutrino mass matrix are crucial for the indirect detection of DM annihilation. However, as will demonstrate in the subsequent sections, there is significant complementarity between the DMID signals, collider experiments, and flavor-violating observables. Each of these aspects is sensitive to different measurements, providing additional insights into neutrino oscillation phenomena through their complementarity. Prior to delving into this discussion, we briefly summarize the constraints on the Yukawa couplings $f_{ij}, g_{ij}$, as well as the cubic coupling~$\mu_{\eta\kappa}$, which are relevant for neutrino mass generation.

For the sake of perturbativity we restrict the Yukawa couplings to be~$f_{ij},g_{ij} < 3$ and the cubic coupling is bounded from above as
\begin{align}
    \mu_{\eta\kappa} \leq \begin{cases}
        \left( 3 \times 2\pi^{2} \right)^{1/4} m_{\eta} &\mathrm{for} \ m_{\kappa} \ll m_{\eta} \\
        \left( 3 \times 6\pi^{2} \right)^{1/4} m_{\eta} &\mathrm{for} \ m_{\kappa} \approx m_{\eta} \\
        \left( 3 \times 24\pi^{2} \right)^{1/4} m_{\eta} &\mathrm{for} \ m_{\kappa} \gg m_{\eta}
    \end{cases} 
\end{align}
due to the requirement that the tree-level quartic couplings must be larger than the effective BSM scalar couplings while respecting perturbativity bounds~\cite{Babu:2002uu}. A more conservative bound on the cubic coupling comes from avoiding a charge-breaking vacuum~\cite{Herrero-Garcia:2014hfa}, manifesting itself in
\begin{align}
    \mu_{\eta\kappa} \lesssim \sqrt{20\pi} \mathrm{max} \left( m_{\eta}, m_{\kappa} \right) \, .
\end{align}
The most relevant experimental constraints come from charged lepton flavor-violating (cLFV) processes, including BSM contributions to the anomalous magnetic dipole moment of the electron and muon, muonium-antimuonium conversion, and lastly, deviations from lepton universality. At the tree level, the presence of the doubly charged scalar introduces new contributions to various lepton flavor-violating processes. Specifically, in trilepton decay processes, these contributions alter the decay width, which can be expressed as
\begin{align}
\frac{\Gamma\left( \ell_{i}^{-} \rightarrow \ell_{j}^{+}\ell_{k}^{-}\ell_{l}^{-}  \right)}{\Gamma\left( \ell_{i}^{-} \rightarrow \ell_{j}^{-} \overline{\nu} \nu \right)} &= \frac{\vert g_{ij} g_{kl}^{*} \vert^{2}}{2 \left( 1+\delta_{kl} \right) G_{F}^{2} m_{\kappa}^{4}} \, .
\end{align}
Moreover, the transition from muonium to antimuonium also experiences a new contribution mediated by the doubly-charged scalar. The corresponding effective coupling is given by
\begin{align}
G_{\mathrm{M}\overline{\mathrm{M}}} = - \frac{g_{ee} g_{\mu \mu}^{*}}{\sqrt{32} m_{\kappa}^{2}} \, .
\end{align}
These tree-level processes yield constraints on the couplings, as summarized in Tab.~\ref{Tab:ZeeBabu_LeptonflavorViolationTreeLevelLimits}.
%%%%%%%%%%%%%%%%%%%%
\begin{table}[t!]
    \centering
    \begin{tabular}{ccc}
        \toprule
        Process & Experiment ($90\%$ C.L.) & Bound  ($90\%$ C.L.)\\
        \midrule
        $\mu^{-} \rightarrow e^{+} e^{-} e^{-}$ & $\mathrm{BR} < 1.0\times 10^{-12}$ & $\left\vert g_{e\mu} g_{ee}^{*} \right\vert/m_{\kappa}^{2} < 2.33 \times 10^{-11} \GeV^{-2}$ \\
        $\tau^{-} \rightarrow e^{+} e^{-} e^{-}$ & $\mathrm{BR} < 2.7\times 10^{-8}$ & $\left\vert g_{e\tau} g_{ee}^{*} \right\vert/m_{\kappa}^{2} < 9.07 \times 10^{-9} \GeV^{-2}$ \\
        $\tau^{-} \rightarrow e^{+} e^{-} \mu^{-}$ & $\mathrm{BR} < 1.8\times 10^{-8}$ & $\left\vert g_{e\tau} g_{e\mu}^{*} \right\vert/m_{\kappa}^{2} < 5.23 \times 10^{-9} \GeV^{-2}$ \\
        $\tau^{-} \rightarrow e^{+} \mu^{-} \mu^{-}$ & $\mathrm{BR} < 1.7\times 10^{-8}$ & $\left\vert g_{e\tau} g_{\mu\mu}^{*} \right\vert/m_{\kappa}^{2} < 7.20 \times 10^{-9} \GeV^{-2}$ \\
        $\tau^{-} \rightarrow \mu^{+} e^{-} e^{-}$ & $\mathrm{BR} < 1.5\times 10^{-8}$ & $\left\vert g_{\mu\tau} g_{ee}^{*} \right\vert/m_{\kappa}^{2} < 6.85 \times 10^{-9} \GeV^{-2}$ \\
        $\tau^{-} \rightarrow \mu^{+} \mu^{-} e^{-}$ & $\mathrm{BR} < 2.7\times 10^{-8}$ & $\left\vert g_{\mu\tau} g_{e\mu}^{*} \right\vert/m_{\kappa}^{2} < 6.50 \times 10^{-9} \GeV^{-2}$ \\
        $\tau^{-} \rightarrow \mu^{+} \mu^{-} \mu^{-}$ & $\mathrm{BR} < 2.1\times 10^{-8}$ & $\left\vert g_{\mu\tau} g_{\mu\mu}^{*} \right\vert/m_{\kappa}^{2} < 8.11 \times 10^{-9} \GeV^{-2}$ \\
        $\mu^{+}e^{-} \rightarrow \mu^{-} e^{+}$ & $G_{\mathrm{M}\overline{\mathrm{M}}} < 0.003\times G_{F}$ & $\left\vert g_{ee} g_{\mu\mu}^{*} \right\vert/m_{\kappa}^{2} < 1.97 \times 10^{-7} \GeV^{-2}$ \\
    \hline
  \end{tabular}
  \caption{Constraints on the Yukawa couplings from lepton flavor-violating processes at tree-level. See Refs.~\cite{SINDRUM:1987nra, Hayasaka:2010np} for the experimental limits.}  \label{Tab:ZeeBabu_LeptonflavorViolationTreeLevelLimits}
\end{table}
%%%%%%%%%%%%%%%%%%%%%%%%%%%%%%%%%%%
Now, at the one-loop level, there are additional contributions to cLFV decays originating from singly and doubly charged scalars. The partial width for cLFV decays at this loop level can be represented as
\begin{align}
    \frac{\Gamma\left( \ell_{i}^{-} \rightarrow \ell_{j}^{-} \gamma \right)}{\Gamma\left( \ell_{i}^{-} \rightarrow \ell_{j}^{-} \overline{\nu} \nu \right)} &= \frac{\alpha}{48\pi G_{F}^{2}} \left[ \left( \frac{\left(f^{\dagger}f\right)_{ij}}{m_{\eta}^{2}} \right)^{2} + 16\left( \frac{\left(g^{\dagger}g\right)_{ij}}{m_{\kappa}^{2}} \right)^{2} \right] \, .
\end{align}
Additionally, there are extra contributions to the anomalous magnetic dipole moment~$a\equiv ( g-2 )/2$ of the electron and muon at this loop level attributed to the charged scalars, and the contribution is given by~\cite{Nebot:2007bc}
\begin{align}
    \delta a_{i} = -\frac{m_{i}^{2}}{24\pi^{2}} \left[ \frac{\left( f^{\dagger} f \right)_{ii}}{m_{\eta}^{2}} + 4\frac{\left( g^{\dagger} g \right)_{ii}}{m_{\kappa}^{2}} \right] \, .
\end{align}
The constraints on the Yukawa couplings resulting from these loop-level cLFV processes and the anomalous magnetic dipole moment of the electron and muon are summarized in Tab.~\ref{Tab:ZeeBabu_LeptonflavorViolationLoopLevelLimits}.
%%%%%%%%%%%%%%%
\begin{table}[b!]
    \centering
    \begin{tabular}{cc}
        \toprule
        Experiment ($90\%$ C.L.) & Bound  ($90\%$ C.L.)\\
        \midrule
        $\mathrm{BR} \left( \mu \rightarrow e\gamma \right) < 4.2 \times 10^{-13}$ & $\frac{\vert f_{e\tau}^{*} f_{\mu\tau} \vert^{2}}{\left(m_{\eta}/\mathrm{GeV} \right)^{4}} + 16 \frac{\vert g_{ee}^{*} g_{e\mu} + g_{e\mu}^{*}g_{\mu\mu} + g_{e\tau}^{*} g_{\mu\tau}\vert^{2}}{\left( m_{\kappa}/\mathrm{GeV} \right)^{4}} < 1.10 \times 10^{-18}$ \\
        $\mathrm{BR} \left( \tau \rightarrow e\gamma \right) < 3.3 \times 10^{-8}$ & $\frac{\vert f_{e\mu}^{*} f_{\mu\tau} \vert^{2}}{\left( m_{\eta}/\mathrm{GeV}\right)^{4}} + 16 \frac{\vert g_{ee}^{*} g_{e\tau} + g_{e\mu}^{*}g_{\mu\tau} + g_{e\tau}^{*} g_{\tau\tau}\vert^{2}}{\left(m_{\kappa}/\mathrm{GeV}\right)^{4}} < 4.85 \times 10^{-13}$ \\
        $\mathrm{BR} \left( \tau \rightarrow \mu\gamma \right) < 4.4 \times 10^{-8}$ & $\frac{\vert f_{e\mu}^{*} f_{e\tau} \vert^{2}}{\left(m_{\eta}/\mathrm{GeV}\right)^{4}} + 16 \frac{\vert g_{e\mu}^{*} g_{e\tau} + g_{\mu\mu}^{*}g_{\mu\tau} + g_{\mu\tau}^{*} g_{\tau\tau}\vert^{2}}{\left(m_{\kappa}/\mathrm{GeV}\right)^{4}} < 6.65 \times 10^{-13}$ \\
        $\delta a_{e} = 2.8 \times 10^{-13}$ & $\frac{\vert f_{e\mu} \vert^{2} + \vert f_{e\tau} \vert^{2}}{\left(m_{\eta}/\mathrm{GeV}\right)^{2}} + 4 \frac{\vert g_{ee} \vert^{2} + \vert g_{e\mu} \vert^{2} + \vert g_{e\tau} \vert^{2}}{\left(m_{\kappa}/\mathrm{GeV}\right)^{2}} < 2.53 \times 10^{-4}$ \\
        $\delta a_{\mu} = 2.61 \times 10^{-9}$ & $\frac{\vert f_{e\mu} \vert^{2} + \vert f_{\mu\tau} \vert^{2}}{\left(m_{\eta}/\mathrm{GeV}\right)^{2}} + 4 \frac{\vert g_{e\mu} \vert^{2} + \vert g_{\mu\mu} \vert^{2} + \vert g_{\mu\tau} \vert^{2}}{\left(m_{\kappa}/\mathrm{GeV}\right)^{2}} < 5.53 \times 10^{-5}$ \\
    \hline
  \end{tabular}
    \caption{Constraints on the Yukawa couplings from lepton flavor-violating processes at loop level and measurements of the anomalous magnetic moments. The experimental limits are quoted from Refs.~\cite{BaBar:2009hkt, MEG:2016leq}.}
    \label{Tab:ZeeBabu_LeptonflavorViolationLoopLevelLimits}
\end{table}
%%%%%%%%%%%%%%%%%%%
Tests of lepton universality also constrain the interactions of singly-charged scalars contributing to $\ell_{i} \rightarrow \ell_{j} \overline{\nu} \nu$. The constraints read~\cite{HFLAV:2022esi}
\begin{align}
    \frac{g_{\tau}}{g_{\mu}} &= \left\vert \frac{1 + \vert f_{e\tau} \vert^{2} v^{2} / m_{\eta}^{2}}{1 + \vert f_{e\mu} \vert^{2} v^{2} / m_{\eta}^{2}} \right\vert = 1.0009 \pm 0.0014 \,,\\
    \frac{g_{\tau}}{g_{e}} &= \left\vert \frac{1 + \vert f_{\mu\tau} \vert^{2} v^{2} / m_{\eta}^{2}}{1 + \vert f_{e\mu} \vert^{2} v^{2} / m_{\eta}^{2}} \right\vert = 1.0027 \pm 0.0014 \,,\\
    \frac{g_{\mu}}{g_{e}} &= \left\vert \frac{1 + \vert f_{\mu\tau} \vert^{2} v^{2} / m_{\eta}^{2}}{1 + \vert f_{e\tau} \vert^{2} v^{2} / m_{\eta}^{2}} \right\vert = 1.0019 \pm 0.0014\,,
\end{align}
and we consider a $3\sigma$ uncertainty range in the analysis. 

The analysis of DM indirect detection will elucidate the intricate interplay between the scalar mediator and various lepton flavors. This relationship determines the expected cosmic ray positron spectrum resulting from DM annihilation within our Galaxy and establishes a connection between neutrino oscillation observables and distinctive flavored signatures of DM. A critical aspect involves scrutinizing the decay processes of singly and doubly charged scalars into leptons, guided by the neutrino mass matrix in the Zee-Babu model. Assuming $m_{\eta}\geq m_{\kappa}$, the doubly charged scalar predominantly decays into same-sign charged leptons, while the singly charged scalar predominantly decays into a charged lepton and neutrino. Understanding the branching ratios of these two BSM scalar fields provides valuable insights into the flavor-specific DM indirect detection signal.

The branching fraction of charged scalars to different lepton flavors can be expressed as 
\begin{align}
    \mathrm{BR} \left( \eta^{+} \rightarrow \ell_{i}^{+} \nu_{j} \right) &= \lv f_{ij} \rv^{2} / \sum_{a\leq b} \lv f_{ab} \rv^{2} \,\\
    \mathrm{BR} \left( \kappa^{++} \rightarrow \ell_{i}^{+} \ell_{j}^{+} \right) &= \frac{\lv g_{ij} \rv^{2}}{1+ \delta_{ij}} / \sum_{a\leq b} \frac{\lv g_{ab} \rv^{2}}{1+ \delta_{ab}}\, .
\end{align}

\begin{figure}[b!]
    \centering
    $$
    \includegraphics[width=0.49\textwidth]{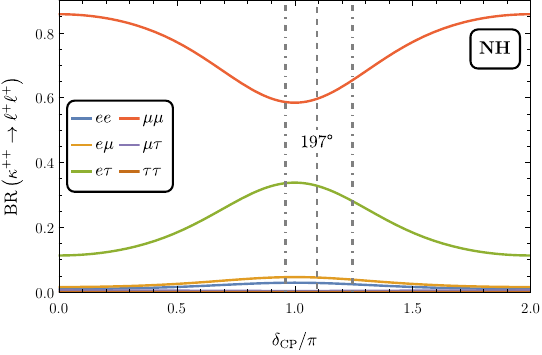}
    \includegraphics[width=0.49\textwidth]{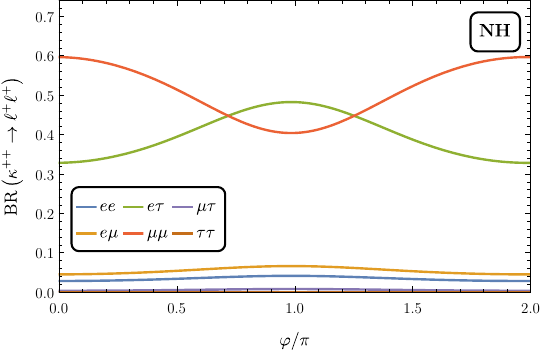}
    $$
    $$
    \includegraphics[width=0.49\textwidth]{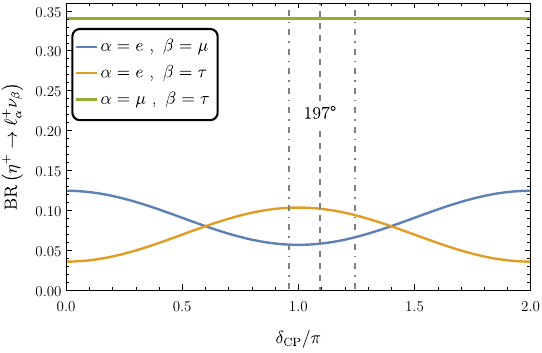}
    $$
    \caption{Branching ratios of doubly charged scalar~$\kappa^{++}$ for the benchmark point BM1~(\emph{top}) and of the singly charged scalar~$\eta^{+}$~(\emph{bottom}) for different lepton flavors. Note that the branching ratio of the singly charged BSM scalar is independent of the scalar mass (as long as its decay channels are open) and the Majorana phase.}
    \label{Fig:ZeeBabu_kappaEta_BR}
\end{figure}

In Fig.~\ref{Fig:ZeeBabu_kappaEta_BR}, we show how the branching ratios of charged scalars vary with different neutrino oscillation parameters, like $\delta_{\rm CP}, \varphi$.
%%%%%%%%%%%%%%%%%%%%%%%%%%%%%%%%%%%%%%%
%%%%%%%%%%%%%%%%%%%%%%%%%%%%%%%%%%%%%%%
By fixing the values of three mixing angles ($\theta_{12}$, $\theta_{23}$, $\theta_{13}$) and two mass differences ($\Delta m_{21}^{2}$, $\Delta m_{3\ell}^{2}$) to their respective central values, it becomes evident that the decay of doubly and singly charged scalars  into specific flavors of leptons is influenced by the CP phases, providing additional insights into neutrino oscillation phenomena. To ensure the comprehensiveness of our investigation, it is imperative to establish specific benchmark points for the Zee-Babu model. This step is essential to guarantee that the benchmark points selected for DM indirect detection analysis and collider analysis remain consistent with neutrino oscillation data, as well as theoretical and experimental constraints. The details of these benchmark points are succinctly presented in Tab.~\ref{Tab:ZeeBabu_Merged_Vertical}.
%%%%%%%%%%%%%%%%%%%%%%%%%%%%%%%%%%%%%%%%%%%
\begin{table}[b!]
    \centering
    \begin{tabular}{llll}
        \toprule
        Benchmark Points & BM1 (min-$\mu$) & BM2 (pure-$\mu$) & BM3 (pure-$\mu$) \\
        \midrule
        $m_{\kappa} \, \left[\mathrm{TeV}\right]$ & 9 & 9 & 1 \\
        $m_{\eta} \, \left[\mathrm{TeV}\right]$ & 29 & 55 & 13 \\
        $\mu_{\eta\kappa} \, \left[\mathrm{TeV}\right]$ & 65 & 99 & 35 \\
        $f_{e\mu}$ & 0.13 & 0.068 & -0.079 \\
        $g_{ee}$ & 0.038 & -0.0014 & 0.0010 \\
        $g_{e\mu}$ & 0.034 & -0.0012 & 0.00051 \\
        $g_{e\tau}$ & -0.092 & 0.00049 &  -0.00066 \\
        $\mathrm{BR}_{ee}$ & 0.042 & $< 10^{-3}$ & $< 10^{-3}$ \\
        $\mathrm{BR}_{e\mu}$ & 0.067 & $< 10^{-3}$ & $< 10^{-3}$ \\
        $\mathrm{BR}_{e\tau}$ & 0.500 & $< 10^{-3}$ & $< 10^{-3}$ \\
        $\mathrm{BR}_{\mu\mu}$ & 0.388 & 0.982 & 0.982 \\
        $\mathrm{BR}_{\mu\tau}$ & 0.007 & 0.018 & 0.005 \\
        $\mathrm{BR}_{\tau\tau}$ & $< 10^{-3}$ & $< 10^{-3}$ & $< 10^{-3}$ \\
        \bottomrule
    \end{tabular}
    \caption{Benchmark points for Zee-Babu model with branching ratios for $\kappa^{++}$~decay for the Dirac CP phase from Tab.~\ref{Tab:NeutrinoOscillationData} and the Majorana phase~$\varphi=\pi$.}
    \label{Tab:ZeeBabu_Merged_Vertical}
\end{table}

We have outlined how neutrino observables influence the interactions between prospective scalar mediators and various lepton flavors within the Zee-Babu model. This analysis is also applicable to other radiative neutrino mass models involving charged scalars that can function as \textit{scalar neutrino portals}, such as the one in Ref.~\cite{Babu:2020hun}, where neutrino masses and mixings emerge at the one-loop level (cf. Fig.~\ref{Fig:NeutrinoMassModels_FeynmanDiagrams}c) and require a quadruplet of scalars.
This is also true for another radiative model, namely Zee model~\cite{Zee:1980ai}, which involves two Higgs doublets~$H_{1,2}:(\mathbf{1},\mathbf{2},1/2)$ and a charged scalar singlet $\eta^{+}: (\mathbf{1},\mathbf{1},1)$, with the $SU(3)_{c} \times SU(2)_{L} \times U(1)_{Y}$ charges given in brackets. If the singly charged scalar responsible for neutrino mass generation acts as a scalar-neutrino-portal for DM, then the Yukawa coupling combination generating the neutrino flavor structure will similarly yield lepton flavor-specific signatures, impacting both cosmic ray positron spectra and collider observations.

\section{Dark Matter phenomenology} \label{Sec:DarkMatter}

As mentioned above, we consider DM to be a complex scalar in this 
work, stabilized by a discrete $\mathbb{Z}_{2}$~symmetry that prevents its decay.
We focus on the WIMP regime, where the coupling of DM to the SM is strong enough to ensure thermalisation in the early Universe.
Annihilation processes keep the DM abundance to its equilibrium abundance as the bath temperature~$T$ drops below the DM mass and the DM abundance gets Boltzmann suppressed.
The relic abundance is finally determined by the freeze-out of DM annihilation processes, i.e.\ the time when the annihilation rate drops below the Hubble rate.
The WIMP relic abundance is inversely proportional to the thermally averaged annihilation cross section~$\sigmav$. To reproduce the observed relic abundance $(\Omega_\mathrm{DM} h^{2})_\mathrm{obs} = 0.1200 \pm 0.0012$~\cite{Planck:2018vyg}, we require (e.g.~\cite{Steigman:2012nb,Bringmann:2020mgx})
\begin{equation}
    \label{eq:sigmavObs}
    \sigmav_{S^{\dagger} S \rightarrow \text{bath}}^{\text{thermal relic}} \simeq 4 \times 10^{-26} \cm^{-3} \second^{-1}\,.
\end{equation}
This requires substantial coupling between the DM scalar and the thermal bath, which can manifest in particle physics processes occurring today in the lab or in the Galaxy.

The simplest possibility of coupling a scalar WIMP to the SM bath is through the Higgs portal, $V \supset \lambda_{SH} (S^{\dagger}S) (H^\dagger H)$.
As the Higgs has substantial coupling to nuclei, $\lambda_{SH}$ is strongly constrained by DM direct detection searches (e.g.~\cite{Cline:2013gha}). In view of current constraints from e.g.~the XENONnT~\cite{XENON:2023cxc} and LZ experiments~\cite{LZ:2022lsv}, Higgs-portal complex scalar WIMPs are only viable for DM masses just below half of the Higgs mass, where annihilation can proceed via resonance and even small values of~$\lambda_{SH}$ can lead to the observed DM relic abundance, or for $m_{S} \gtrsim 3 \TeV$.
The present work focuses on the case where other mediators than the SM Higgs are responsible for the DM-SM interaction, and we set Higgs portal coupling~$\lambda_{SH}=0$ in the following. 
There is a question that may arise at this moment, that how the thermal equilibrium of DM can still be maintained with negligible Higgs portal interaction. This is possible if the particles that the DM interacts with are also in equilibrium, which is evidently true for the doubly/singly charged scalars as mediators originating from neutrino mass models.

We have already discussed in the previous two sections that such interactions of DM may provide a crucial hint to neutrino mass generation mechanisms and provide distinctive features via leptonic signals in indirect or collider searches. The portal terms for the DM in both cases then read
\begin{equation}
    \label{Eq:Vportal}
    V_\mathrm{portal}^{\Delta} = \lambda_{S\Delta} S^{\dagger}S \mathrm{Tr} \left( \Delta^{\dagger} \Delta \right)\,,
    \qquad
    V_\mathrm{portal}^{\mathrm{ZB}} = \lambda_{S\eta} \eta^{-}\eta^{+} S^{\dagger}S + \lambda_{S\kappa} \kappa^{--}\kappa^{++} S^{\dagger}S\,.
\end{equation}
In the case where $m_\mathrm{DM} > m_\mathrm{mediator}$, these lead to the thermally averaged annihilation cross section
\begin{equation}
    \label{eq:sigmavScalar}
    \sigmav_{S^{\dagger} S \rightarrow \text{mediator pair}}^{\text{non-relativistic}} = \frac{\lambda^2 \sqrt{1-m_{\text{mediator}}^2/m_S^2}}{32 \pi m_S^2} \,,
\end{equation}
where $\lambda^{2}=3 \lambda_{S\Delta}^{2}$ corresponds to the case of the mass-degenerate triplet (using $v_{\Delta} \ll v_{H}$), and $\lambda^2= \lambda_{S\kappa}^2,\lambda_{S\eta}^2$ to annihilation into the doubly or singly charged scalars in the Zee Babu model. To reproduce the observed relic abundance in the case $m_{S} \ll m_{\mathrm{mediator}}$, we require $\lambda \sim 0.6 \left( m_{S}/\TeV \right)$.

\begin{figure}[b!]
    \centering
    \includegraphics[width=0.7\textwidth]{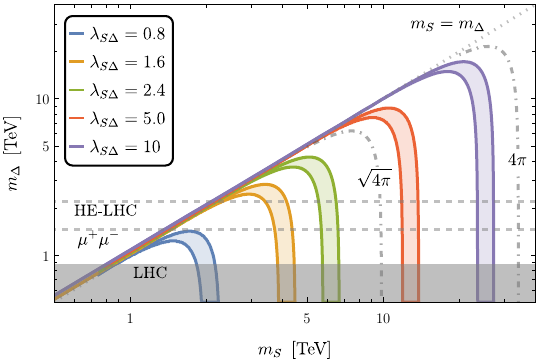}
    \caption{Parameter space for a DM relic abundance $0.75 \leq \Omega_{\mathrm{DM}} h^{2}/(\Omega_{\mathrm{DM}} h^{2})_{\mathrm{obs}} \leq 1$ with the portal coupling~$\lambda_{S\Delta}$ and degenerate masses $m_{\Delta}$ of the fields in the $\Delta$ triplet. The BSM vev is set to~$v_{\Delta}=10^{-5}\GeV$. The shaded parameter space is subject to present LHC bounds. The dashed lines correspond to projected bounds of a future $\mu^{+}\mu^{-}$ collider and the HE-LHC (see text for details).}
    \label{Fig:DM_MassPlane}
\end{figure}

For an accurate calculation of the DM relic abundance, we make use of the public \micrOMEGAs code~\cite{Belanger:2018ccd}.
Fig.~\ref{Fig:DM_MassPlane} shows values of the $\Delta$ mass (for the fully degenerate case) that reproduce the observed relic abundance for different values of the portal coupling in the type-II seesaw plus singlet scenario.
%%%%%%%%%%%%%%%%%%%%%%%%%%%%%%%%%%%%%%%%%%%%%
%%%%%%%%%%%%%%%%%%%%%%%%%%%%%%%%%%%%%%%%%%%%%
Equivalent results in the Zee-Babu scenario can be obtained by re-scaling the coupling appropriately, i.e.\ $\lambda_{S\kappa} = \sqrt{3} \lambda_{S\Delta}$ if DM annihilates only into $\kappa^{++}\kappa^{--}$ or \ $\lambda_{S\eta} = \sqrt{3} \lambda_{S\Delta}$ if DM annihilates only into $\eta^{+}\eta^{-}$.
From the figure, it is clear that for $m_{\Delta} \ll m_{S}$ the relic abundance depends only on $\lambda_{S\Delta}$ and $m_S$, leading to an upper limit on the DM mass, above which the DM abundance cannot be sufficiently depleted with perturbative couplings.
At $m_{\Delta} \gtrsim m_{S}$, the relic abundance depends sensitively on $m_{\Delta}$ as the annihilation process becomes Boltzmann suppressed.
The lower limit on the viable DM mass range then results from collider bounds on the mediator mass.
Assuming decays solely into electrically charged leptons, the strongest bounds by the ATLAS collaboration excludes $m_{\Delta^{++}}\equiv m_{\Delta}\lesssim 870\GeV$~\cite{ATLAS:2017xqs} with present LHC data.
Fig.~\ref{Fig:DM_MassPlane} also shows projections for the HE-LHC at~$\sqrt{s}=27\TeV$ and integrated luminosity~$\mathcal{L}_{\mathrm{int}} = 15\ab^{-1}$ at~$m_{\Delta^{++}}< 2.2\TeV$~\cite{Padhan:2019jlc} and the reach of a future $\mu^{+}\mu^{-}$-collider with $\sqrt{s} = 3\TeV$ and ~$\mathcal{L}_{\mathrm{int}}=1000\fb^{-1}$ at~$m_{\Delta^{++}} \lesssim 1.45\TeV$~\cite{Maharathy:2023dtp}. 
Much of the parameter space amenable to perturbative analysis may be probed by future colliders, and we explore flavorful implications in detail in Section~\ref{Sec:Collider}.

Therefore, scalar WIMP DM coupled through leptophilic scalars that arise in simple neutrino mass models are viable in the mass range $\mathcal{O}(1-10)\TeV$ (apart from the Higgs resonance regions), limited towards low masses by collider constraints on the mediator particles and towards high masses by perturbativity.

We have further verified that DMDD constraints on the BSM-mediated scenarios arise at one-loop level, and they do not constrain the parameter space as long as the couplings~$\lambda_{H\Delta}$ ($\lambda_{H\eta},\lambda_{H\kappa}$) are in the perturbative regime. The dominant signature expected of DM in the present scenario is then its annihilation in our Galaxy today. In the following, we explore these DMID implications and ask whether the flavored annihilation products can carry information about neutrino massses and be used to discern between the different models.

%%%%%%%%%%%%%%%%%%%%%%%%%%%%%%%%%%%%%%%%%%%%%%%%%%%%%%%%%%%%%%%
\section{Dark matter indirect detection and flavor dependence}
The dominant signature expected of DM in the present scenario is the annihilation process $S^\dagger S \rightarrow M^{\dagger} M$ with the subsequent decay of the mediator ($M$) into leptons $M \rightarrow \ell_{i} \ell_{j} (\ell_{i} \nu_{j})$.
In this section, we investigate how the flavorful couplings of the neutrino-portal scalars introduced in Sections~\ref{Sec:DMinTypeII} and~\ref{Sec:DMinZeeBabu} impact the DMID signal.

\begin{figure}[b!]
    \centering
    \includegraphics[width=0.49\textwidth]{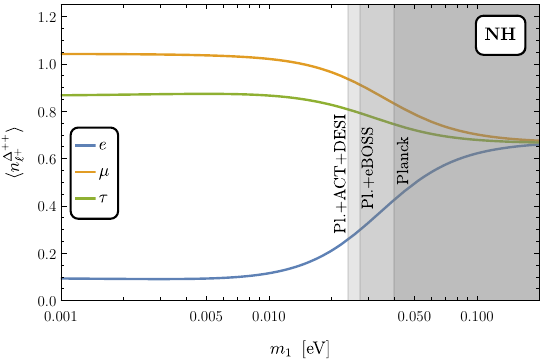}
    \includegraphics[width=0.49\textwidth]{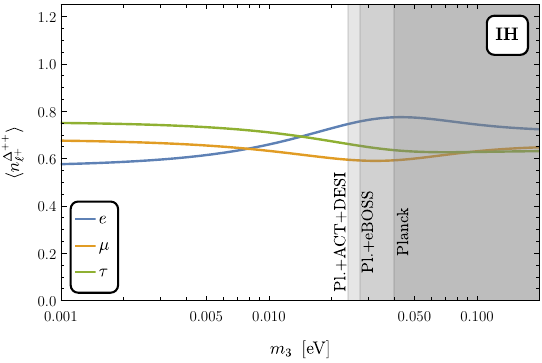}\\
    \includegraphics[width=0.49\textwidth]{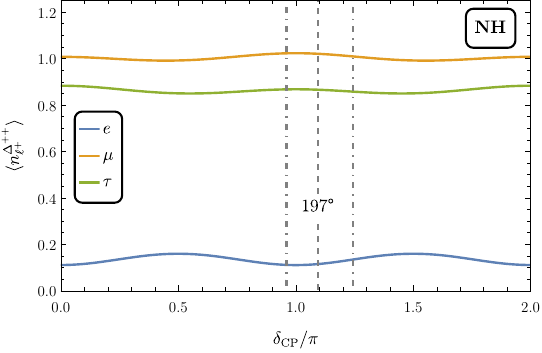}
    \includegraphics[width=0.49\textwidth]{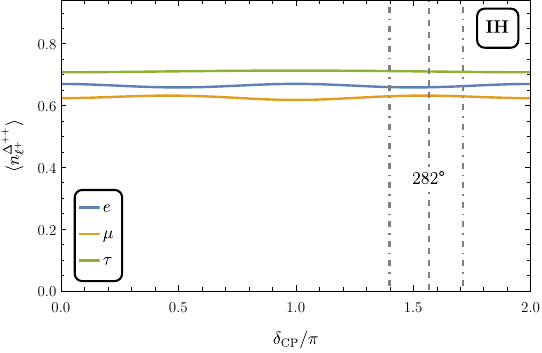}\\
    \includegraphics[width=0.49\textwidth]{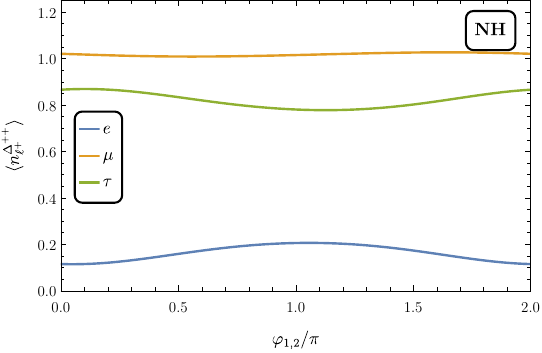}
    \includegraphics[width=0.49\textwidth]{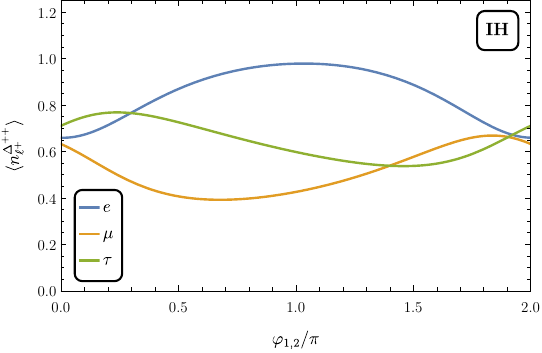}
    \caption{Mean number of lepton species $\braket{n_{\ell^{+}}^{\Delta^{++}}}$ per $\Delta^{++}$ decay in the type-II seesaw scenario. The parameter assignment is the same as for Fig.~\ref{Fig:TypeII_Delta_BR}.}
    \label{Fig:TypeII_MeanLeptonNumber}
\end{figure}

\begin{figure}[b!]
    \centering
    \includegraphics[width=0.49\textwidth]{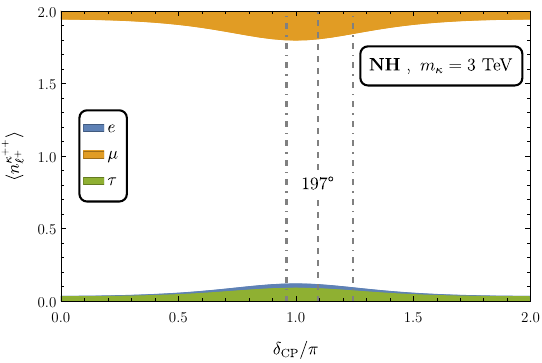}
    \includegraphics[width=0.49\textwidth]{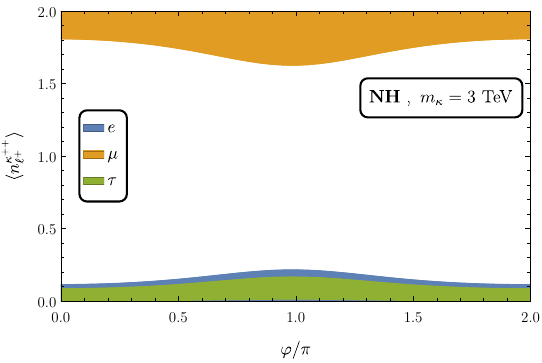} \\
    \includegraphics[width=0.49\textwidth]{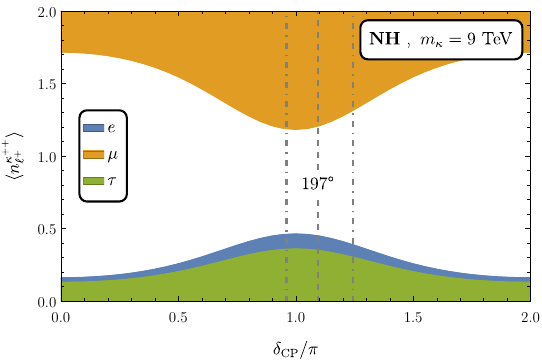}
    \includegraphics[width=0.49\textwidth]{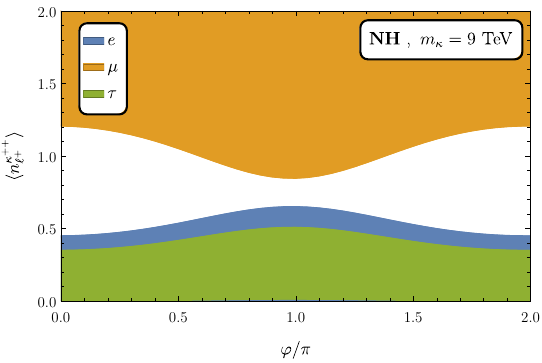}\\
    \includegraphics[width=0.49\textwidth]{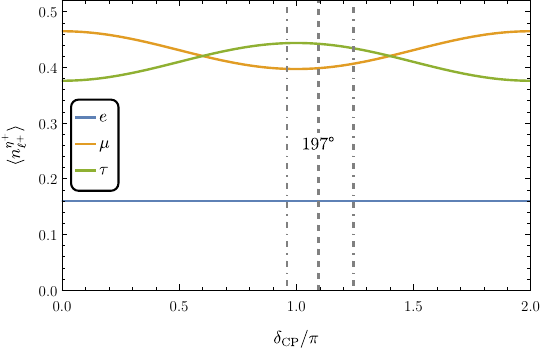}    
    \caption{Mean number of leptons with the Dirac CP phase as given in Tab.~\ref{Tab:NeutrinoOscillationData} and the Majorana phase~$\varphi=0$ if not stated otherwise. \emph{Top rows:} Bands of possible mean numbers of leptons in the decay of~$\kappa^{++}$. The free Yukawa couplings in~$f,g$ as well as the cubic coupling~$\mu_{\eta\kappa}$ are sampled. The parameter points satisfy the theoretical and experimental constraints. The plots in the left column contain information on the central value of~$\delta_{\mathrm{CP}}$ (dashed) as well as its $1\sigma$ uncertainty range (dot-dashed). \emph{Bottom:} Mean numbers of leptons in the decay of~$\eta^{+}$.
    }
    \label{fig:ZeeBabu_kappa_MeanNumber}
\end{figure}

The mediator couplings to the different leptons determine the relative production rate of primary leptons $e,\mu,\tau$.
A signal from DM annihilation in the cosmos will average over many annihilation events.
Hence, DMID is not sensitive to the individual mediator branching ratios, but only to the average number $\braket{n_{\ell^{+}}^{M}}$ of $\ell^{+}$ produced in one decay of a mediator particle~$M$.
With~$\ell\neq \ell^\prime, \ell^{\prime\prime}$ and the mediator $\Delta^{++}$ (or $\kappa^{++}$), this reads
\begin{align}
    \braket{n_{\ell^{+}}^{\kappa^{++}}}\equiv 2 \times \mathrm{BR} \left( \kappa^{++} \rightarrow \ell^{+}\ell^{+} \right) + \mathrm{BR} \left( \kappa^{++} \rightarrow \ell^{+}\ell^{\prime +} \right) +\mathrm{BR} \left( \kappa^{++} \rightarrow \ell^{+}\ell^{\prime \prime +} \right)\, ,
\end{align}
whereas for the singly charged $\eta^{+}$ (or $\Delta^{+}$) it is
\begin{align}
    \braket{n_{\ell^{+}}^{\eta^{+}}}\equiv \mathrm{BR} \left( \eta^{+} \rightarrow \ell^{+}\nu_{\ell} \right) + \mathrm{BR} \left( \eta^{+} \rightarrow \ell^{+}\nu_{\ell^{\prime}} \right) +\mathrm{BR} \left( \eta^{+} \rightarrow \ell^{+}\nu_{\ell^{\prime\prime}} \right)\, .
\end{align}
The averaged production of primary leptons $\braket{n_{\ell^{+}}^{M}}$ is shown in 
Fig.~\ref{Fig:TypeII_MeanLeptonNumber} for the type-II seesaw scenario and in
Fig.~\ref{fig:ZeeBabu_kappa_MeanNumber} for the Zee-Babu model.
%%%%%%%%%%%%%%%%%%%%%%%%%%%%%%%%%%%%%%%%
%%%%%%%%%%%%%%%%%%%%%%%%%%%%%%%%%%%%%%%%%%
Summing the different contributions leads to loss of information compared to the individual branching ratios (Figs.~\ref{Fig:TypeII_Delta_BR},\ref{Fig:ZeeBabu_kappaEta_BR}), to which collider studies are in principle sensitive.
Nevertheless, it is clear that sensitivity to the different scenarios, as well as to neutrino mass ordering and potentially neutrino phases is retained.
In particular, in the type-II seesaw scenario, NH and IH ordering of neutrino masses results in a clear difference in the number of primary positrons vs.\ muons or taus produced. The dependence on the phases is mild, except for the IH case, where the channels differ substantially as a function of $\varphi_{1,2}$.
In the Zee-Babu case, there are two possible mediators, $\kappa^{++}$ and $\eta^+$.
The values of $\braket{n_{\ell^{+}}^{\kappa^{++},\eta^{+}}}$ allow to distinguish which of the scalars the DM dominantly annihilates into.
In the case of $\kappa^{++}$, the decay proceeds mostly into $\mu^{+}$, with increasing room for other decay channels as the mass scale is raised and constraints on couplings in the other channels are relaxed, which allows for some sensitivity to the neutrino phases.
In the case of $\eta^{+}$, there is an $\mathcal{O}(10\%)$ variation in the $\mu^{+}$ and $\tau^{+}$ channels that will be hard to discriminate.
To assess whether any of these variations translate to discernible differences in the DMID signal, we consider the benchmark points given in Tab.~\ref{Tab:MeanNumberBenchmarks}.
%%%%%%%%%%%%%%%%%%%%%%%%%%%%%%%%%%%%%%%%%%%%%%
\begin{table}[b!]
    \centering
    \small
    \begin{tabular}{lccccccccc}
        \toprule
        $\mathrm{BM}$ & $\mathrm{BR}_{ee}$ & $\mathrm{BR}_{e\mu}$ & $\mathrm{BR}_{e\tau}$ & $\mathrm{BR}_{\mu\mu}$ & $\mathrm{BR}_{\mu\tau}$ & $\mathrm{BR}_{\tau\tau}$ & $\braket{n_{e^{\pm}}}$ & $\braket{n_{\mu^{\pm}}}$ & $\braket{n_{\tau^{\pm}}}$\\
        \midrule
        SS-II NH & $0.03$ & $0.05$ & $0.02$ & $0.22$ & $0.54$ & $0.16$ & $0.12$ & $1.02$ & $0.86$\\
        SS-II NH $\varphi$ & $\sim 0$ & $0.19$ & $0.01$ & $0.13$ & $0.56$ & $0.10$ & $0.21$ & $1.02$ & $0.78$\\
        SS-II IH & $0.29$ & $0.04$ & $0.03$ & $0.09$ & $0.42$ & $0.13$ & $0.66$ & $0.63$ & $0.71$\\
        SS-II IH $\varphi$ & $0.02$ & $0.38$ & $0.56$ & $0.02$ & $0.01$ & $0.01$ & $0.98$ & $0.43$ & $0.60$\\
        \\
        ZB-$\kappa$ pure-$\mu$ & $\sim 0$ & $\sim 0$ & $\sim 0$ & $0.98$ & $0.02$ & $\sim 0$ & $\sim 0$ & $1.98$ & $0.02$\\
        ZB-$\kappa$ min-$\mu$ & $0.04$ & $0.07$ & $0.50$ & $0.39$ & $0.01$ & $\sim 0$ & $0.65$ & $0.85$ & $0.50$\\
        ZB-$\eta$ & $0$ & $0.06$ & $0.10$ & $0$ & $0.34$ & $0$ & $0.16$ & $0.40$ & $0.44$\\
        \hline
    \end{tabular}
    \caption{Benchmark points for branching ratios and mean number of leptons in different neutrino mass models.
    The type-II SS model benchmark points correspond to the two neutrino mass hierarchies and either~$\varphi_{1,2} = 0$ or $\varphi_{1,2} = \pi$ (labelled with~$\varphi$ for the latter), while $m_{1,3} = 0.01\eV$ and the Dirac CP phase is fixed to the value in Tab.~\ref{Tab:NeutrinoOscillationData}; corresponding model parameters can be obtained via Eq.~(\ref{Eq:SSIIyukawas}).
    The Zee-Babu model benchmark points correspond to NH, $m_1=0$, $\varphi = \pi$, and Dirac CP phase as in Tab.~\ref{Tab:NeutrinoOscillationData}. For example model parameters see Tab.~\ref{Tab:ZeeBabu_Merged_Vertical}.}
    \label{Tab:MeanNumberBenchmarks}
\end{table}
%%%%%%%%%%%%%%%%%%%%%%%%%%%%%%%%%%%%%%%

The actual DMID signal of interest is the cosmic ray positron flux arising from present-day DM annihilation  in our Galaxy. Cosmic ray detectors measure the flux $\dd\Phi_{e^\pm}/\dd E = v_{e^\pm} n/4\pi$, where $n=\dd N/\dd E \, \dd^{3}x$ is the cosmic ray density.
Cosmic ray positron propagation in the Galaxy can be described by a diffusion-loss equation (e.g.~\cite{Atoyan:1995ux})
\begin{equation}
    \frac{\partial n}{\partial t} 
    - \nabla \left( D(E,\vec x) \nabla n\right)
    - \frac{\partial}{\partial E} \left( b(E,\vec x) n\right)
    = Q(E,\vec x)\,,
    \label{eqn:PropagationEqn}
\end{equation}
where $D(E,\vec x)$ is a diffusion coefficient, $b(E,\vec x)$ the effective energy loss rate\footnote{
It has been pointed out recently that the discontinuous energy loss of multi-TeV positrons experiencing inverse Compton scattering can enhance the dark matter signal close to the endpoint~\cite{John:2023ulx} (see also Ref.~\cite{Zdziarski1989ApJ...342.1108Z}) compared to the continuous approximation. For the energies considered here, the enhancement is expected to be $\mathcal{O}(0.1)$.}
and $Q(E,\vec x)$ the cosmic ray source term.
The source term due to DM annihilation is given by
\begin{align}
    Q\left( E_{0},\vec{x}_{0} \right) =
    \frac{1}{4}\frac{\braket{\sigma v} \rho_{\mathrm{DM}}^{2} \left( \vec{x}_{0} \right)}{m_{\mathrm{DM}}^{2}}
    \frac{\dd N_{\mathrm{ann}}}{\dd E}\left( E_0 \right) \, ,
    \label{eqn:CRsourceTerm}
\end{align}
where the factor $1/4$ accounts for the fact that only $S^{\dagger}S$-pairs can annihilate in the present scenario, $\rho_{\mathrm{DM}}$ is the galactic DM density, $\sigmav$ the thermally averaged annihilation cross section given in Eq.~(\ref{eq:sigmavObs}), and lastly~$\dd N_\mathrm{ann} /\dd E$ the positron spectrum per DM annihilation event.

To assess the impact of flavorful DM annihilation on the positron flux observable at Earth, we first compare positron spectra $\dd N_\mathrm{ann} /\dd E$ at the DM annihilation site and then investigate two scenarios for DM annihilation: the guaranteed signal from annihilation in the smooth DM halo of the Milky Way~(MW), as well as an optimistic scenario in which a nearby dense subhalo leads to a large additional cosmic ray flux.

\subsection{Cosmic ray source term}
As discussed in the previous sections, every DM annihilation event produces an average number $\braket{n_{\ell^{+}}}$ of lepton $\ell^{+}$. The initially produced charged leptons quickly decay to particles that are stable on galactic scales.
Cosmic ray positrons are the most sensitive probe of $e$/$\mu$-philic DM, with some sensitivity to $\tau$-philic DM~\cite{John:2021ugy,DiMauro:2021qcf}. For this reason, we focus on cosmic ray positrons.
If a signal is detected, correlated signals in other channels, particularly in gamma rays, and in other astrophysical environments are expected and will serve to confirm its DM nature.

In the present scenario, DM annihilates into a pair of mediators, which subsequently decay to leptons that themselves result in positrons.
The positron spectrum per annihilation event can be calculated by appropriately boosting the mediator decay products:
\begin{align}
    \frac{\dd N^M_{\mathrm{ann}}}{\dd E} =
    \int_{E^0_\mathrm{min}}^{E^0_\mathrm{max}} dE_0
    \;
    \left(\sum_M
    \frac{\sigmav_M}{\sigmav_\mathrm{tot}}
    \langle n^M_{\ell^{+}} \rangle
    \frac{\dd N^M_{\ell^{+}}}{\dd E_0}
    \right)
    \times \left( E_0 \frac{m_S}{m_M} \sqrt{1-\frac{m_M^2}{m_S^2}} \right)^{-1}\,,
    \label{Eq::dndeAnn}
\end{align}
where~$E^0_\mathrm{min}= E \left( \frac{m_S}{m_M} \left(1+\sqrt{1-\frac{m_M^2}{m_S^2}}\right) \right)^{-1}$ and
$E^0_\mathrm{max}= E \left( \frac{m_S}{m_M} \left(1-\sqrt{1-\frac{m_M^2}{m_S^2}}\right) \right)^{-1}$.
Here $\frac{\dd N^M_{\ell^{+}}}{\dd E_0}$ is the normalised spectrum of positrons resulting from the decay of the mediator into initial lepton $\ell^{+} \in \{e^+,\mu^+,\tau^+\}$ in the mediator rest frame and $\langle n^M_{\ell^{+}} \rangle$ is the average number of initial lepton $\ell^{+}$ produced per mediator decay.
$\frac{\sigmav_M}{\sigmav_\mathrm{tot}}$ is the ratio of annihilations proceeding into mediator $M$ compared to the total annihilation rate.
In the type-II seesaw model, we consider the components $\Delta^{++},\Delta^+,\Delta_{R,I}^{0}$ to be mass degenerate, for concreteness.
In the Zee-Babu model, we assume that either the doubly charged scalar or the singly charged scalar is much heavier than the DM candidate and does not contribute to DM annihilation.
We obtain the decay spectrum of the mediator using the tabulated results from Refs.~\cite{Cirelli:2010xx,Ciafaloni:2010ti} that include contributions from electroweak final state radiation, $\dd N_{\ell^{+}}/\dd E_0 = \left.\dd N_{\ell^{+}}^\mathrm{PPPC}/\dd E_0\right|_{m_\mathrm{DM}=m_M/2}$.

\begin{figure}
    \centering
    \includegraphics[width=0.49\textwidth]{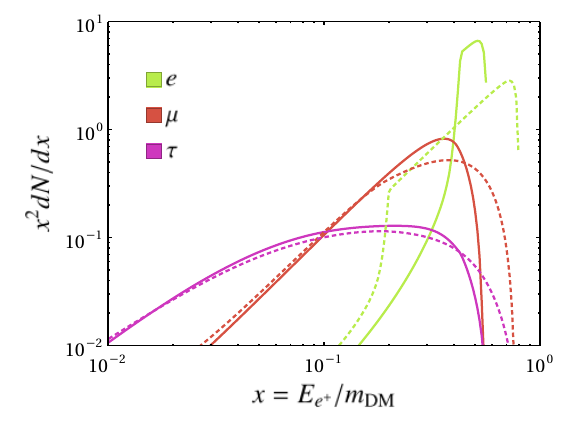}
    \includegraphics[width=0.49\textwidth]{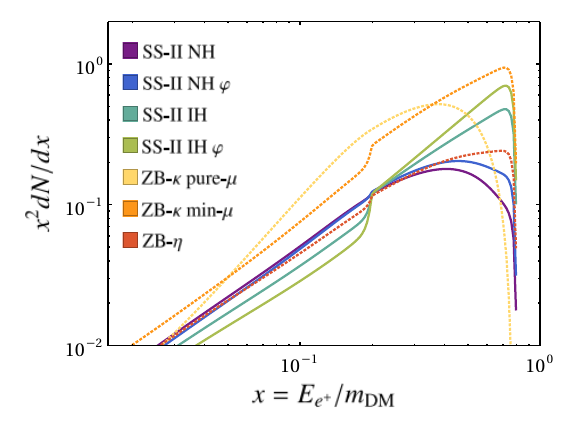}
    \caption{Positron injection spectra per DM annihilation event. Lines: $m_\mathrm{DM}/m_\mathrm{mediator} = 1/0.99 \TeV$ benchmark, dashed: $10/8 \TeV$.
    \emph{Left:} Positron spectra assuming the mediator decays purely into a single lepton flavor.
    \emph{Right:} Positron spectra corresponding to the neutrino mass model benchmark points given in Tab.~\ref{Tab:MeanNumberBenchmarks} for $m_\mathrm{DM}/m_\mathrm{mediator} = 10/8 \TeV$.
    }
    \label{fig:injectionSpectra}
\end{figure}

The resulting spectra are shown in Fig.~\ref{fig:injectionSpectra} for two benchmark sets of DM and mediator mass, for different annihilation channels.
One can see that the electron flavor results in the most peaked positron spectra, while annihilation into muons or tauons results in flatter peaks at lower energies.
The larger the boost of the mediators, the broader the peaks become. Yet the spectra from the different initial lepton flavors remain distinguishable also in the more boosted benchmark (dashed).
The neutrino mass model benchmarks of Tab.~\ref{Tab:MeanNumberBenchmarks} correspond to different mixtures of the primary $e/\mu/\tau$ contributions. This is shown in the right panel of Fig.~\ref{fig:injectionSpectra}.
The different branching fractions in particular to the peaked electron channel result in distinguishable spectra.
Annihilation purely to doubly charged mediators in the $\mathrm{ZB-}\kappa$ case results in larger normalisation than to the singly charged mediators in the $\mathrm{ZB-}\eta$ case or the mixed production of $\Delta^{++,+,0}$ in the type-II seesaw case. Propagation from the site of DM annihilation to the detector close to Earth will tend to smooth these spectra, as we will see in the following.

\subsection{Annihilation in the smooth Milky Way halo}
To estimate the guaranteed Galactic DM annihilation signal, we assume the DM profile of the Galaxy to follow an NFW~\cite{Navarro:1996gj} profile
\begin{align}
    \rho_{\mathrm{NFW}} \left( r \right) = \rho_{\mathrm{gal}} \frac{r_{\mathrm{gal}}}{r} \left( 1 + \frac{r}{r_{\mathrm{gal}}} \right)^{-2}\,,
\end{align}
with the characteristic scale $r_{\mathrm{gal}}=24.42\kpc$ and $\rho_{\mathrm{gal}}=0.184\GeV\cm^{-3}$, resulting in a local DM density of $\rho_\odot=0.3\GeV/\cm^3$~\cite{Cirelli:2010xx}.
We calculate the cosmic ray flux resulting from DM annihilation throughout the Galaxy using pre-calculated propagation functions from the PPPC~\cite{Cirelli:2010xx,Buch:2015iya}, using the ``MAX'' propagation scenario with magnetic field configuration ``MF1''.
This simple treatment is justified given the small (as we will see) expected signal and the large uncertainty in the DM halo and Galactic magnetic field.

Fig.~\ref{fig:fluxPlots} shows the expected positron fluxes for two different DM/mediator mass benchmarks and different neutrino parameters.
Compared to the injection spectra in Fig.~\ref{fig:injectionSpectra}, the peaks are smoothed out, but the drop towards their endpoints still carries some flavor information. Concretely, the larger the DM annihilation fraction into electrons, the more pronounced is the drop-off of the spectra towards the kinematic endpoint.
This may in principle enable e.g.\ distinguishing the type-II seesaw IH scenario from the NH or Zee-Babu scenarios, assuming the astrophysical background is sufficiently smooth.
However, the overall flux is low -- more than an order of magnitude below current AMS-02 data~\cite{AMS:2021nhj,Maurin:2013lwa} and possibly beyond the reach of AMS-100~\cite{Schael:2019lvx} for TeV-scale DM masses.

As the annihilation rate scales as $\rho^2$, there can be significant enhancement by DM substructure on scales smaller than the positron diffusion scale. Such an astrophysical ``boost factor'' is expected in the present WIMP DM scenario and can lead to 1-2 orders-of-magnitude larger cosmic ray fluxes (e.g. Refs.~\cite{Delos:2022bhp,Stucker:2023rjr}), bringing the expected flux closer to future detectability.
Even larger boost factors have been invoked in the literature, but are subject to stringent gamma-ray constraints (e.g.~\cite{Lin:2014vja}).

The overall emission morphology in the Milky Way would however be similarly extended as in the smooth-halo case shown here, and we hence expect similar spectral shapes of the resulting fluxes.

\subsection{Signal from a nearby subhalo}

In N-body simulations, DM halos of galaxies like the Milky Way are found to contain an abundance of subhalos~\cite{Springel:2008cc}. If such a DM substructure is close to the Earth, it would act as an additional source of DM-annihilation associated cosmic rays and could potentially lead to enhanced signals (see e.g.\ Ref.~\cite{Cumberbatch:2006tq}).
This is particularly interesting for the case of positrons, where a more local source leads to a positron flux that more closely resembles the injected flux than the propagation-smoothened contribution from the smooth MW halo considered in the previous section (see e.g.\ Refs.~\cite{Yuan:2017ysv,Jin:2017qcv}).

In the case of a nearby source, the diffusion and loss terms in the propagation Eq.~(\ref{eqn:PropagationEqn}) can be approximated as spatially homogeneous, and the local flux can be determined in a Greens function approach (e.g.~\cite{Atoyan:1995ux,Kuhlen:2009is}):
\begin{align}
    n\left( \vec{x}, E \right) = \frac{1}{b\left( E \right)} \int \dd^{3} x_{0} \int_{E}^{\infty} \dd E_{0} \, Q\left( \vec{x}_{0},E_{0} \right) \, G\left( \vec{x}-\vec{x}_{0},E,E_{0} \right)\,,
\end{align}
with
\begin{align}
    G\left( \vec{x}-\vec{x}_{0},E,E_{0} \right) = \left( 4\pi \lambda \left( E, E_{0} \right)\right)^{-3/2} \exp{\left( - \frac{\left( \vec{x}-\vec{x}_{0} \right)^{2}}{4 \lambda \left( E, E_{0} \right)} \right)} \,.
\end{align}
Here $\sqrt{\lambda(E,E_0)}$ is a diffusion scale, given by
\begin{align}
    \lambda \left( E, E_{0} \right) = \int_{E}^{E_{0}} \dd E^{\prime} \, \frac{D \left( E^{\prime} \right)}{b \left( E^{\prime} \right)} \, . 
\end{align}
For $D(E)$ and $b(E)$, we use the values used in Ref.~\cite{Cirelli:2010xx} for the ``MAX'', ``MF1'' configuration.

We consider the case of a point-like subhalo at distance $d=0.1\kpc$ from Earth.
This is well within the diffusion scale even close to the maximum energy ($\sqrt{\lambda}\gtrsim 0.2 \kpc$ at 90\% of the injected energy for all energies considered), which justifies the point-like approximation and leads to minimal washout of the spectral information.
The source term is given by $Q(\vec x, E) = \delta(\vec x - \vec x_\mathrm{subhalo}) \mathcal{L} \, \dd N/\dd E$, with subhalo luminosity $\mathcal{L} = \int \mathop{d^3x} \rho_\mathrm{subhalo}^2$.
We take $\mathcal{L}=10^{62} \, \mathrm{GeV^2}/\mathrm{cm}^3$ for concreteness.
Assuming a power-law subhalo density profile $\rho_\mathrm{subhalo}(r) \propto r^{-\alpha}$, this luminosity corresponds to a subhalo of mass $\sim 10^{4-5} M_\odot$ with $1<\alpha<1.5$, where we have truncated the subhalo at the tidal radius~\cite{vandenBosch:2017ynq} appropriate to the solar position in the Galaxy.

\begin{figure}
    \centering
    \includegraphics[width=0.49\textwidth]{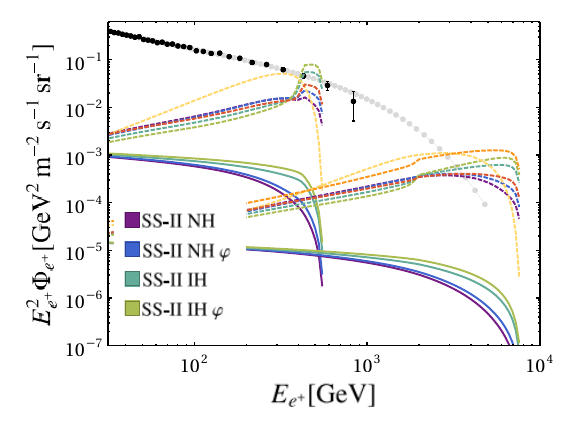}
    \includegraphics[width=0.49\textwidth]{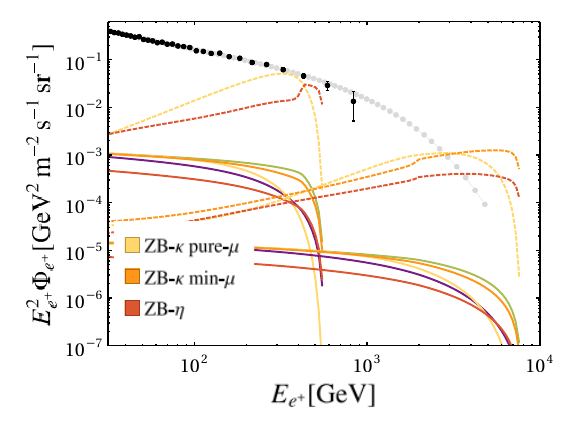}
    \caption{Local cosmic ray positron fluxes, split into two panels for readability, contrasted with AMS-02 data (black) and an AMS-100 forecast (gray). Lines: guaranteed signal from the smooth MW halo. Dashed: nearby subhalo at 0.1 kpc distance with luminosity $10^{62} \mathrm{GeV^2}/\mathrm{cm}^3$.
    Results for the mass spectra $M_\mathrm{DM}/m_\mathrm{mediator} = 1/0.99,\; 10/8 \TeV$ are shown (with ZB-$\kappa$ min-$\mu$ only allowed for the heavier benchmark).
    The named neutrino mass benchmarks are given in Tab.~\ref{Tab:MeanNumberBenchmarks}.
    }
    \label{fig:fluxPlots}
\end{figure}

Fig.~\ref{fig:fluxPlots} shows the expected positron fluxes from a nearby subhalo as dashed lines for the same DM and neutrino benchmarks as considered in the previous subsection.
Depending on the produced primary lepton flavor composition, the spectra are sharply peaked and the different neutrino scenarios result in distinctive spectral shapes.
The normalisation of the spectra depends linearly on the subhalo luminosity and distance.
For $m_\mathrm{DM} = 1\TeV$, the benchmark subhalo could have already shown up in present data,
while a less luminous subhalo may be detected by future experiments.
For $m_\mathrm{DM} = 10\TeV$, the benchmark subhalo could generate a clear signal at future experiments, with the potential to discriminate between the different neutrino scenarios.

\subsection{Discussion}

In Fig.~\ref{fig:fluxPlots} we compare expected positron fluxes with current data from AMS-02~\cite{AMS:2021nhj,Maurin:2013lwa} (black). As there is no indication of a DM signal in the current data~\cite{John:2021ugy}, we also show a forecast for the proposed AMS-100 detector (which assumes an extrapolation of the smooth spectrum observed by AMS-02)~\cite{Schael:2019lvx}.

Without a boost from substructure, the positron flux from DM annihilation in the present scenario is too small to be detected in present data, or with a more powerful future experiment.
However, small-scale substructure in DM microhalos~\cite{Ishiyama:2010es,Delos:2022bhp,Stucker:2023rjr} may enhance the generic WIMP expectation for the Galactic DM annihilation rate by an order of magnitude, which could bring the Galactic signal closer to detectability.

In the case of a nearby subhalo, the positron flux from DM annihilation could be greatly enhanced, especially at high energies.
This would have promising implications for a future experiment like AMS-100, which could observe the DM-annihilation induced peak in the positron spectrum. Its spectral shape would allow to make inferences on neutrino oscillation parameters.
In particular, it could distinguish between scenarios where the doubly charged mediator couples predominantly to electrons versus muons or taus, as the former results in more peaked positron spectra.  

The probability of the presence of a subhalo like the one shown in Fig.~\ref{fig:fluxPlots} in a MW-like Galaxy at distance $\leq 0.1\kpc$ from a position at the solar radius has been estimated in Ref.~\cite{Jin:2017qcv} based on the Aquarius simulation~\cite{Springel:2008cc} to lie around $10^{-3}$, with probability scaling roughly inversely proportional to the subhalo luminosity.

Note that the identification of DM-induced features in the positron flux depends on the smoothness or subdominance of astrophysical backgrounds. While present data indicate a smooth flux amenable to DM analyses~\cite{John:2021ugy}, effects of distinct nearby positron sources may result in spectral irregularities \cite{Cholis:2021kqk,Krommydas:2022loe,Lv:2024khg} (see however~\cite{John:2022asa}). This would hamper the inference of neutrino properties when the DM-induced flux does not clearly dominate over backgrounds.

The inclusion of additional channels next to positrons may further strengthen the prospects of learning about neutrino mass generation through DMID in the doubly-charged-mediator scenario. In particular, hadronic decays of primary taus result in gamma rays, which may offer a complimentary probe of leptophilic DM annihilation.

\section{Collider complementarity}
\label{Sec:Collider}
%%%%%%%%%%%%%%%%%%%%%%%%%%%%%%%%%%%%%%%%%%%%%%%
\begin{figure}[htb!]
    \centering
    \includegraphics[width=0.32\textwidth]{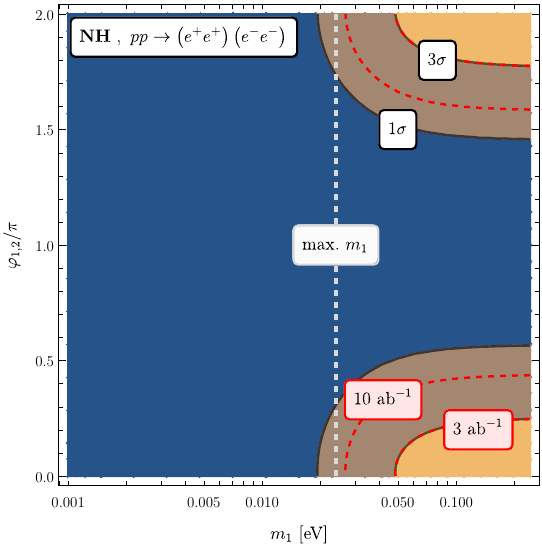}
    \includegraphics[width=0.32\textwidth]{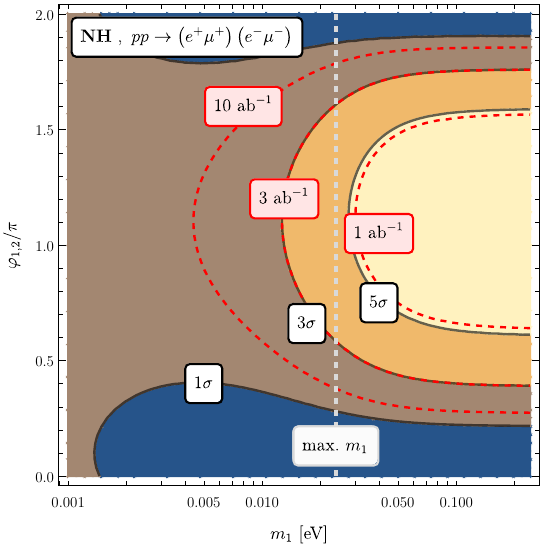}
    \includegraphics[width=0.32\textwidth]{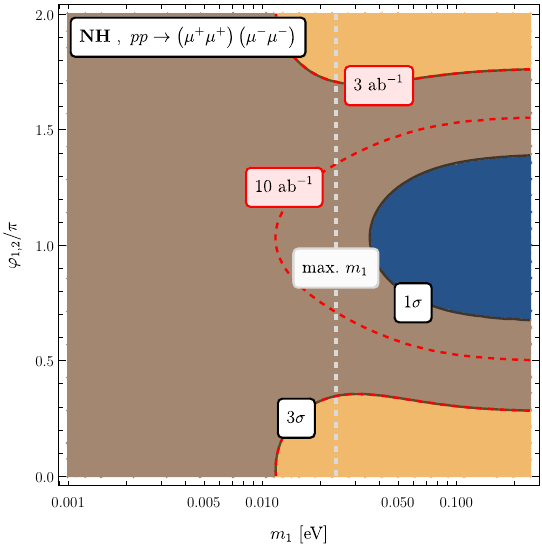}\\
    \includegraphics[width=0.32\textwidth]{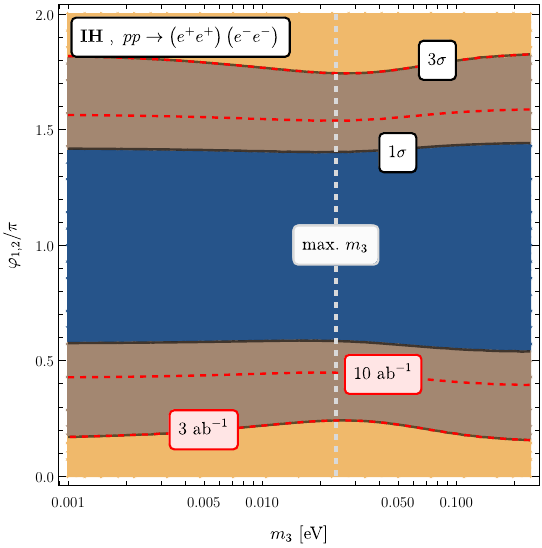}
    \includegraphics[width=0.32\textwidth]{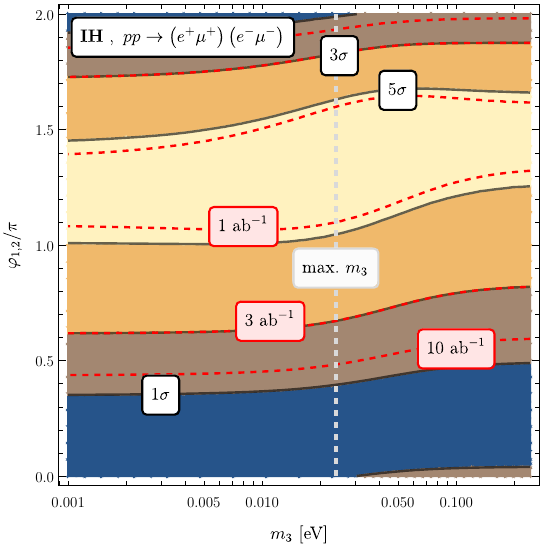}
    \includegraphics[width=0.32\textwidth]{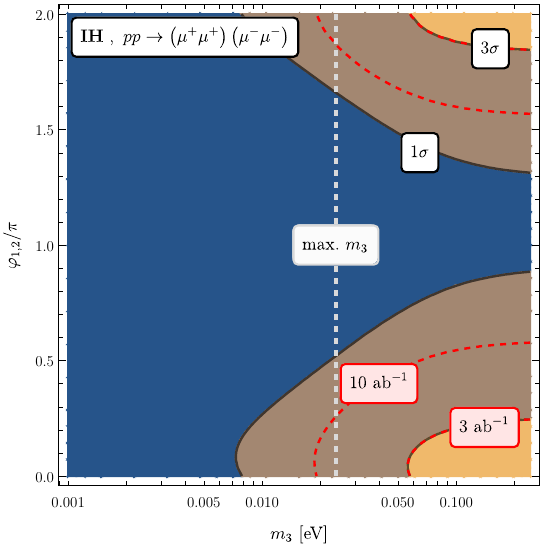}
    \caption{Sensitivities at HL-LHC for neutrino oscillation parameters with respect to the neutrino mass scale~$m_{1,3}$ and the Majorana phase~$\varphi$ (we assume~$\varphi_{1} = \varphi_{2} \equiv \varphi$ and the Dirac CP phase from Tab.~\ref{Tab:NeutrinoOscillationData}) in the type-II seesaw model for a fixed mediator mass $m_{\Delta} = 1\TeV$. The contours indicate the statistical significance at $3\,\mathrm{ab}^{-1}$ integrated luminosity and the red, dashed lines show the \(3\sigma\) contours for different integrated luminosities. The straight vertical line indicates the upper bound for the lightest neutrino mass.}
    \label{Fig:Collider_TypeII_neutrinoMassVSMajPhase}
\end{figure}
%%%%%%%%%%%%%%%%%%%%%%%%%%%%%%%%%%%%%%%%%%%%%
In this section, we investigate the collider implications of DM models considered here and highlight the complementary nature of flavor-specific signals. We outline how neutrino observables influence the interactions between the potential scalar mediators and various lepton flavors, impacting collider searches. Our analysis centers on the signatures of doubly charged scalars, as they produce a pair of same-sign leptons, which are particularly promising discovery channels with nearly zero SM background.

By fixing the three mixing angles ($\theta_{12}$, $\theta_{23}$, $\theta_{13}$) and two mass differences ($\Delta m_{21}^{2}$, $\Delta m_{3\ell}^{2}$) to their central values from Tab.~\ref{Tab:NeutrinoOscillationData}, it is evident that the decay of doubly charged scalars $\Delta^{++}$ into specific lepton flavors is affected by the CP phases ($\delta_{\mathrm{CP}}$, $\varphi_{1}$, $\varphi_{2}$) and the lightest neutrino mass. Upon production via $s$-channel $Z/\gamma$ exchange through the Drell-Yan process, the expected decay modes of the doubly charged scalar into different lepton flavors vary according to the neutrino oscillation parameters, in particular the lightest neutrino mass and CP phases (see Fig.~\ref{Fig:TypeII_Delta_BR}). In the following, we calculate the impact of this on the expected sensitivities at the High-Luminosity LHC (HL-LHC). 
%%%%%%%%%%%%%%%%%%%%%%%%%%%%%%%%%%%%%%%%%%%%%%%
\begin{figure}[htb!]
    \centering
    \includegraphics[width=0.32\textwidth]{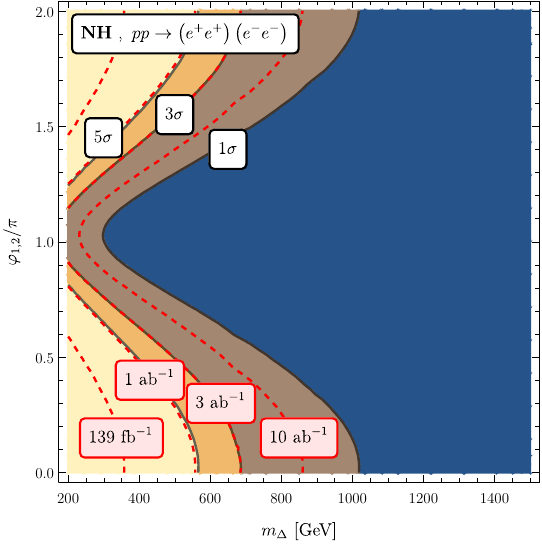}
    \includegraphics[width=0.32\textwidth]{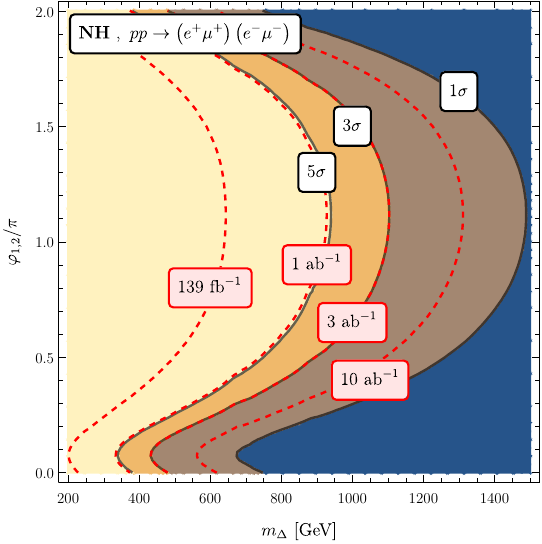}
    \includegraphics[width=0.32\textwidth]{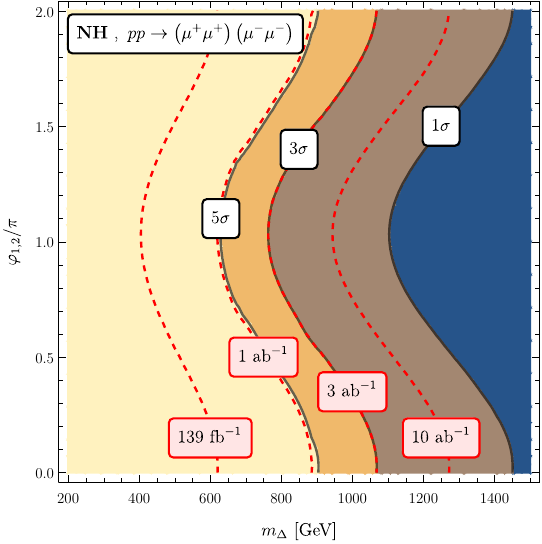}\\
    \includegraphics[width=0.32\textwidth]{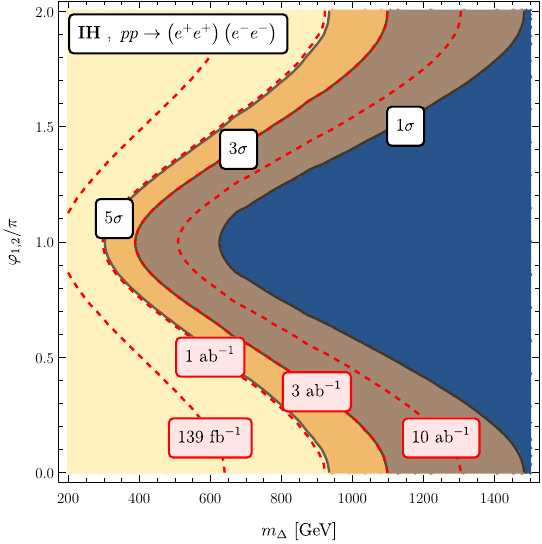}
    \includegraphics[width=0.32\textwidth]{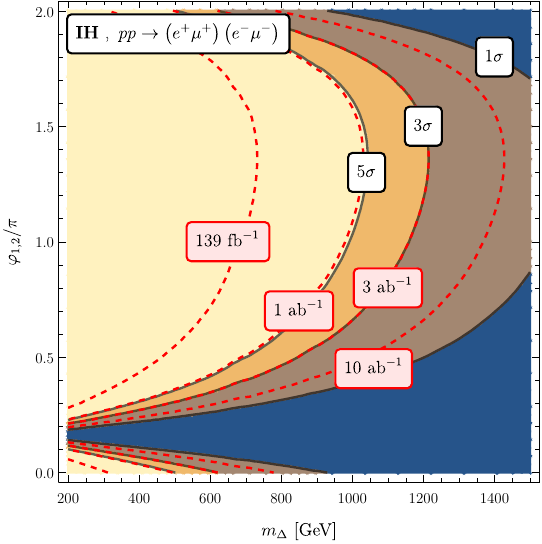}
    \includegraphics[width=0.32\textwidth]{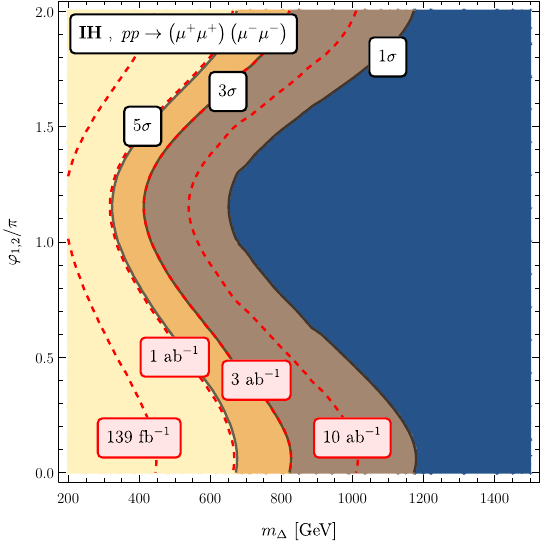}
    \caption{Sensitivities of doubly charged scalars at HL-LHC as a function of the mediator mass~$m_{\Delta}$ and Majorana phase~$\varphi$ in the type-II seesaw model. The Dirac CP phase is taken from Tab.~\ref{Tab:NeutrinoOscillationData} and the neutrino mass scale is set to~$m_{0} = 0.02\eV$. The contours indicate the statistical significance at $3\,\mathrm{ab}^{-1}$ integrated luminosity and the red dashed lines show the $3\sigma$ contours for different integrated luminosities.}
    \label{Fig:Collider_TypeII_mediatorMassVSMajPhase}
\end{figure}
%%%%%%%%%%%%%%%%%%%%%%%%%%%%%%%%%%%%%%%%%%%%%%%
After incorporating our model file into the {\tt FeynRules} package~\cite{Alloul:2013bka}, we compute the cross-section for the Drell–Yan plus photon fusion pair production mode of $H^{++}$ at the 14 TeV LHC using the Monte Carlo event generator {\tt MadGraph5aMC@NLO}~\cite{Alwall:2014hca}. To maximize signal efficiency, we employ the following acceptance criteria: (a) $p_{T}(\ell) > 15\GeV$, (b) $\vert \eta(\ell) \vert < 2.5$, and (c) a veto on any opposite-sign dilepton pair invariant mass near the $Z$ boson mass: $\vert M(\ell^{+}\ell^{-}) - M_{Z} \vert > 15\GeV$. Next, we analyze the discovery reach of $\Delta^{++}$ at the HL-LHC in the four-lepton signal\footnote{Note that such charged scalars can also be long-lived in nature, leading to large ionization energy loss in collider experiments as well as catalyzing fusion processes in light nuclei, potentially impacting energy production \cite{Akhmedov:2024rvp}.} from the decays $\Delta^{++}\rightarrow \ell^{+}\ell^{+}$. Applying the aforementioned cuts results in signal events with negligible SM background. 
Even in the absence of background, charge misidentification and the misclassification of jets as leptons can result in a very minor background. For example, $t\Bar{t}$ events can produce a negligible same-sign dilepton background.
However, reconstructing the invariant mass for same-sign dileptons produces a sharp peak at $m_{\Delta^{++}}$ without any SM background. We compute the predicted number of events as
\begin{align}
    N_{\mathrm{events}} = \sigma_{\mathrm{prod}} \times \mathrm{BR}\left( \Delta^{++} \rightarrow \ell_{a}^{+} \ell_{b}^{+} \right) \times \mathrm{BR}\left( \Delta^{--} \rightarrow \ell_{c}^{-} \ell_{d}^{-} \right) \times \mathcal{L}_{\mathrm{int}}\,,
\end{align}
with the integrated luminosity~$\mathcal{L}_{\mathrm{int}}$ and the significance can then be calculated via
\begin{align}
    \mathcal{S} \equiv N_{\mathrm{events}}/\sqrt{N_{\mathrm{events}} + N_{\mathrm{bkg}}} \ .
\end{align}

\begin{figure}[htb!]
    \centering
    \includegraphics[width=0.32\textwidth]{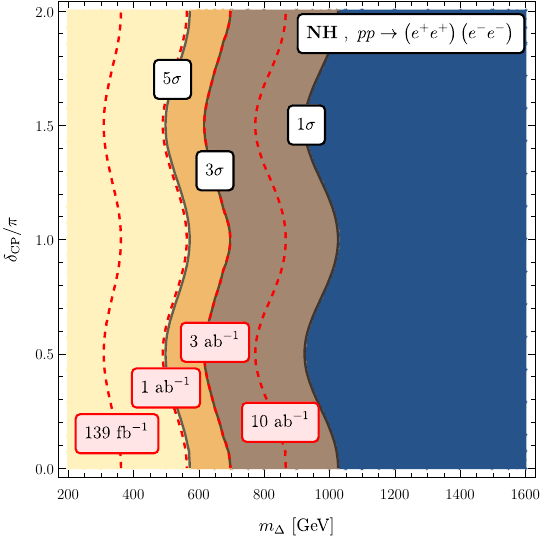}
    \includegraphics[width=0.32\textwidth]{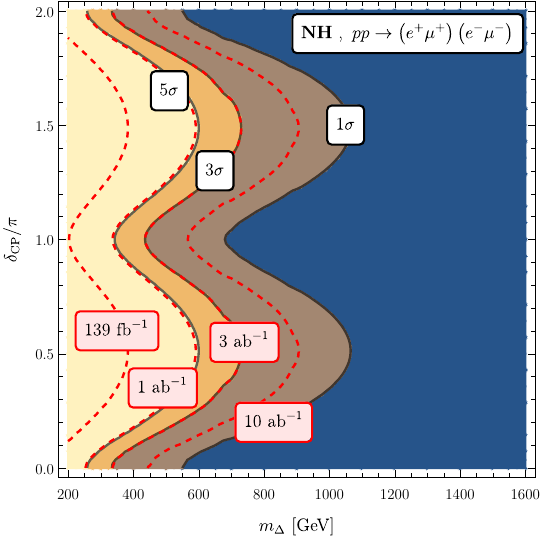}
    \includegraphics[width=0.32\textwidth]{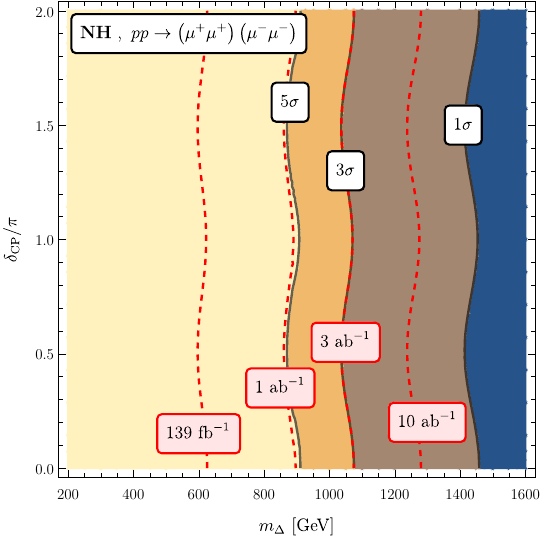}\\
    \includegraphics[width=0.32\textwidth]{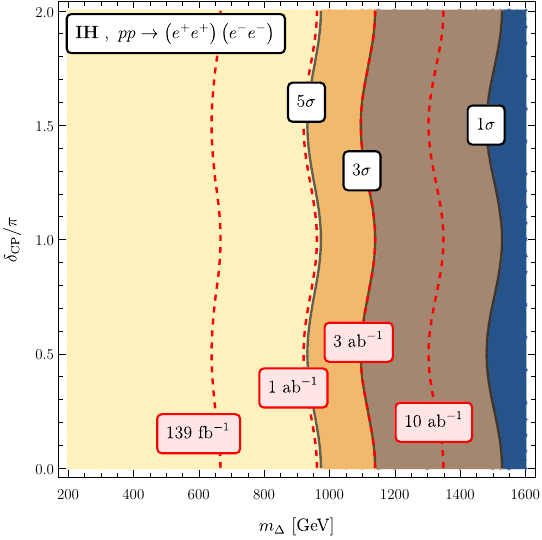}
    \includegraphics[width=0.32\textwidth]{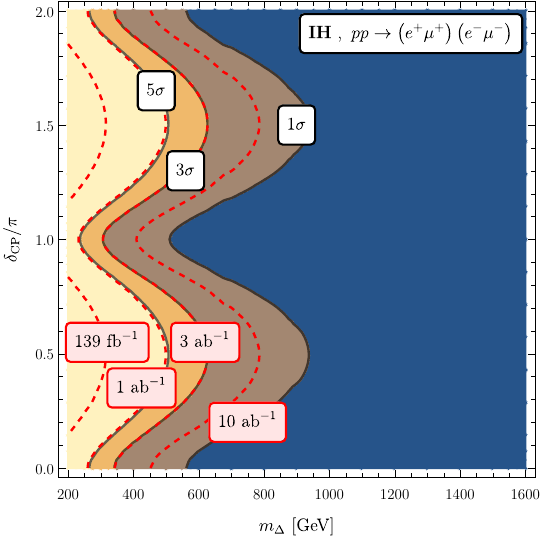}
    \includegraphics[width=0.32\textwidth]{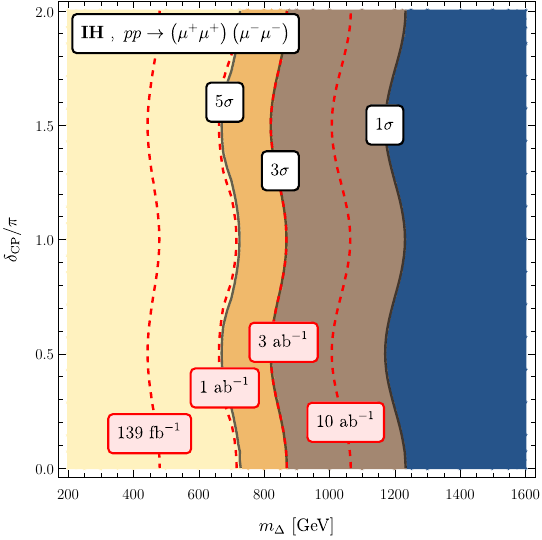}
    \caption{Sensitivities of doubly charged scalars at HL-LHC as a function of the mediator mass~$m_{\Delta}$ and Dirac CP phase~$\delta_{\mathrm{CP}}$ in the type-II seesaw model with the neutrino mass scale~$m_{0} = 0.02\eV$ and the Majorana phase~$\varphi=0$. The contours indicate the statistical significance at $3\,\mathrm{ab}^{-1}$ integrated luminosity and the red dashed lines show the $3\sigma$ contours for different integrated luminosities.}
    \label{Fig:Collider_TypeII_mediatorMassVSDiracPhase}
\end{figure}
%%%%%%%%%%%%%%%%%%%%%%%%%%%%%%%%%%%%%%%%%%%%%%%

Fig.~\ref{Fig:Collider_TypeII_neutrinoMassVSMajPhase} shows the sensitivity to the lightest neutrino mass $m_{0}$ and Majorana phase $\varphi$, assuming $m_{\Delta^{++}}=1\TeV$. By examining the collider signals for the processes $pp \rightarrow \Delta^{++}\Delta^{--} \rightarrow e^{+} e^{+} e^{-} e^{-}/\mu^{+} \mu^{+} \mu^{-} \mu^{-}/e^{+} \mu^{+} e^{-} \mu^{-}$, we find that there is substantial potential to obtain complementary information on neutrino oscillation parameters at the HL-LHC. Fig.~\ref{Fig:Collider_TypeII_mediatorMassVSMajPhase} illustrates the sensitivities of doubly charged scalars at the HL-LHC as a function of the Majorana phase~$\varphi$ in the type-II seesaw model for various integrated luminosities, while Fig.~\ref{Fig:Collider_TypeII_mediatorMassVSDiracPhase} shows the sensitivities of doubly charged scalars at the HL-LHC as a function of the Dirac phase $\delta_{\mathrm{CP}}$ in the type-II seesaw model. It is evident that at the HL-LHC with an integrated luminosity of $10~\mathrm{ab}^{-1}$, a doubly charged scalar with a mass of $1.5\TeV$ can be probed, contingent on different neutrino oscillation parameters. Tau leptons were not considered in the final state, as the identification efficiency for electrons and muons is significantly higher than that for taus at colliders. 

\begin{figure}[htb!]
    \centering
    \includegraphics[width=0.49\textwidth]{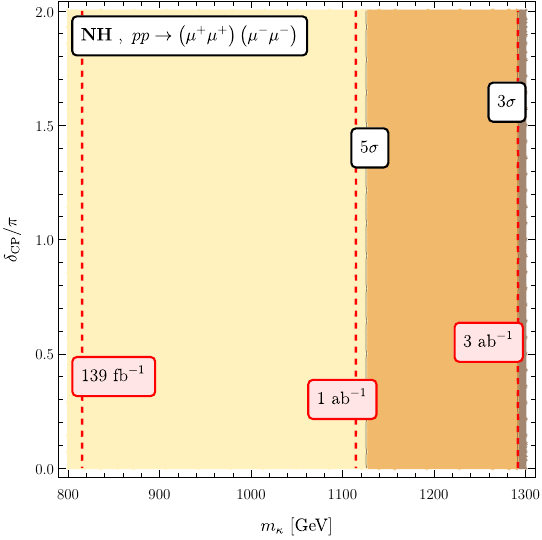}
    \includegraphics[width=0.49\textwidth]{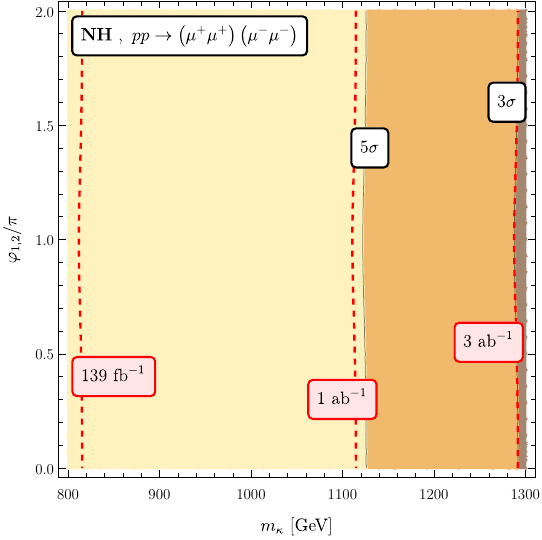}
    \caption{Sensitivities of doubly charged scalars at HL-LHC as a function of the neutrino CP phases~$\delta_{\mathrm{CP}}$, $\varphi_{1,2}\equiv \varphi$, and the mediator mass~$m_{\kappa}$ in the Zee-Babu model for BM3 from Tab.~\ref{Tab:ZeeBabu_Merged_Vertical}. The Dirac CP phase is given by Tab.~\ref{Tab:NeutrinoOscillationData} and the Majorana phases~$\varphi=0$ if not stated otherwise. The contours indicate the statistical significance at $3\,\mathrm{ab}^{-1}$ integrated luminosity, and the red dashed lines show the $3\sigma$ contours for different integrated luminosities. 
    }
\label{Fig:Collider_ZeeBabu_mediatorMassVSDiracPhase}
\end{figure}

Next, we analyze the collider implications of a doubly charged scalar within the Zee-Babu model. Unlike the type-II seesaw scenario, where the doubly charged scalar is an SU(2)$_L$ triplet, the Zee-Babu model involves an SU(2)$_L$ singlet, leading to a different production rate. Fig.~\ref{Fig:Collider_ZeeBabu_mediatorMassVSDiracPhase} presents the estimated sensitivity of the doubly charged scalar at the HL-LHC depending on different CP phases.

In the Zee-Babu model, the lightest neutrino mass is predicted to be zero. Thus, the same-sign dilepton signature at the collider is sensitive to the other neutrino oscillation parameters. By fixing the three mixing angles ($\theta_{12}$, $\theta_{23}$, $\theta_{13}$) and two mass differences ($\Delta m_{21}^{2}$, $\Delta m_{3\ell}^{2}$) to their central values from Tab.~\ref{Tab:NeutrinoOscillationData}, it becomes evident that the decay of doubly charged scalars $\kappa^{++}$ into specific lepton flavors is influenced by the CP phases ($\delta_{\mathrm{CP}}$, $\varphi_{1}$, $\varphi_{2}$), as shown in Fig.~\ref{Fig:Collider_ZeeBabu_mediatorMassVSDiracPhase}.
It is quite important to note that the benchmark points or regions of interest in the Zee-Babu model indicate that the only sensitive signal is \(pp \rightarrow \kappa^{++}\kappa^{--} \rightarrow \mu^{+} \mu^{+} \mu^{-} \mu^{-}\).
This sensitivity arises because the doubly charged scalar almost exclusively decays into pairs of muons within the specified region. Comparing the two models with identical neutrino oscillation parameters clearly demonstrates distinct predictions for the doubly charged scalar scenario.

Similarly, one can investigate the signals of singly charged scalars, triply charged scalars, or other multi-charged scalars at the collider, depending on the different neutrino mass models. In these models, the Yukawa coupling responsible for the decay of such charged scalars also plays a role in neutrino mass generation. However, the study of other charged scalars at the collider is beyond the scope of this paper and is left for future research.

\section{Conclusions}

We investigate the flavor-specific signals of leptophilic scalar DM within a variety of neutrino mass models. In these frameworks, DM signals are directly linked to neutrino oscillation data, offering complementary insights into the neutrino mass hierarchy and CP phases even in the absence of a flavor-specific portal to DM, provided Higgs portal interactions are subdominant. We explore the type-II seesaw and Zee-Babu models in detail, examining the correlation between the flavor-specific signatures of DM and neutrino oscillation data. Our investigation includes the indirect signatures of leptophilic DM, with a particular focus on the cosmic ray electron/positron flux spectrum resulting from DM pair annihilation in the Galactic halo or a nearby DM subhalo. In particular, in the presence of a nearby subhalo, the positron flux from DM could be enhanced, so that future experiments like AMS-100 could observe the DM-annihilation induced peak. The spectral shape could distinguish between scenarios where the doubly charged mediator couples predominantly to electrons, muons or taus, with the former resulting in more peaked positron spectra. Furthermore, we delve into correlated lepton-specific signals at collider experiments that are sensitive to neutrino oscillation data. Doubly charged scalar mediators leading to same-sign dilepton signals provide such opportunities, which can reach discovery limit at HL-LHC and also distinguish between type-II seesaw and Zee-Babu models. Such complementarity can enhance our understanding of the interaction between DM and neutrinos and provides opportunities for experimental verification through cosmic ray observations and collider experiments.

\section*{Acknowledgements}
SJ wishes to acknowledge the Center for Theoretical Underground Physics and Related Areas (CETUP*) and the Institute for Underground Science at SURF for their warm hospitality and for providing a stimulating environment where part of this work was done. SB acknowledges the invitation and hospitality at MPIK Heidelberg, during PASCOS 2022, where the work was planned. 
\bibliographystyle{JHEP}    
\bibliography{Neutrino}

\end{document}